\documentclass[a4paper,12pt]{article}
\pdfoutput=1 

\usepackage{jheppub} 
\usepackage[export]{adjustbox}
\usepackage[utf8]{inputenc}
\usepackage{multirow}

\usepackage{slashed}
\usepackage{bm}

\usepackage{fancyvrb}
\usepackage{fvextra}

\usepackage{comment}
\usepackage[normalem]{ulem}

\usepackage{tikz}
\usepackage{tkz-euclide}
\usetikzlibrary{shapes.misc}
\usetikzlibrary{decorations.pathmorphing}	
\tikzset{
    v/.style={decorate, decoration={snake, segment length=3mm, amplitude=0.75mm}, draw},
    f/.style={draw=black, postaction={decorate},
        decoration={markings,mark=at position .6 with {\arrow[very thick]{latex}}}},
    fb/.style={draw=black, postaction={decorate},
        decoration={markings,mark=at position .4 with {\arrowreversed[very thick]{latex}}}},
    fnar/.style={draw=black},
    g/.style={decorate, draw=black,
        decoration={coil,amplitude=3pt, segment length=3.5pt}},
    s/.style={dashed,draw=black, postaction={decorate},
        decoration={markings,mark=at position .55 with {\arrow[very thick]{latex}}}},
    sb/.style={dashed,draw=black, postaction={decorate},
        decoration={markings,mark=at position .55 with {\arrowreversed[draw=black,very thick]{latex}}}},
    snar/.style={dashed,draw=black,line width =1.25pt},
    cross/.style={cross out, draw=black, minimum size=2*(#1-\pgflinewidth), inner sep=0pt, outer sep=0pt},
cross/.default={3pt},
}
\usepackage{pifont}

\graphicspath{{Figures/}} 

\newcommand{\eqn}[1]{Eq.~(\ref{#1})}

\def\be{\begin{equation}}
\def\ee{\end{equation}}
\newcommand{\ba}{\begin{array}}
\newcommand{\ea}{\end{array}}
\def\bV{\begin{Verbatim}[breaklines=true, breakanywhere=true]}
\def\eV{\end{Verbatim}}

\def\amtwo{a_{\mu}}
\def\damtwo{\Delta a_{\mu}}
\def\amhvp{a_{\mu}^{\rm HVP}}
\def\amhvplo{a_{\mu}^{\rm LO, \rm HVP}}
\def\Mv{m_V}
\def\shad{\sigma_{\rm had}}

\def\Lee{\mathcal{L}_{e^+e^-}}

\def\Mfv{M_{V \phi}}

\def \lsim{\mathrel{\vcenter
     {\hbox{$<$}\nointerlineskip\hbox{$\sim$}}}}
\def \gsim{\mathrel{\vcenter
     {\hbox{$>$}\nointerlineskip\hbox{$\sim$}}}}

\newcommand{\amc}{{\sc MadGraph5}\_a{\sc MC@NLO}}
\newcommand{\fr}{{\sc Feyn\-Rules}}

\newcommand*{\INFNFR}{Istituto Nazionale di Fisica Nucleare, Laboratori Nazionali di Frascati, C.P. 13, 00044 Frascati, Italy}
\newcommand*{\IPII}{Institut de Physique des 2 Infinis de Lyon (IP2I),
UMR5822, CNRS/IN2P3, F-69622 Villeurbanne Cedex, France}

\title{The muon g-2 anomaly confronts new physics in $e^\pm$ and  $\mu^\pm$ final states scattering}

\author[a,b]{Luc Darm\'e,}
\author[b]{Giovanni Grilli di Cortona,}
\author[b]{and Enrico Nardi}

\affiliation[a]{\IPII}
\affiliation[b]{\INFNFR}

\emailAdd{l.darme@ip2i.in2p3.fr}
\emailAdd{ grillidc@lnf.infn.it}
\emailAdd{Enrico.Nardi@lnf.infn.it}

\abstract{
The 4.2$\sigma$ discrepancy between the standard model prediction for the muon  anomalous magnetic moment $a_\mu$ and the experimental result is accompanied by other anomalies.  A crucial input for the prediction is the hadronic vacuum polarization $a_\mu^{\rm HVP}$  inferred from \mbox{$\sigma_{\rm had} =\sigma(e^+e^- \to\,$hadrons)} data.
However, the two most accurate determinations  of  $\sigma_{\rm had}$ from KLOE and BaBar disagree by almost 3$\,\sigma$. Additionally, the combined data-driven result disagrees with the  most precise lattice determination of $a_{\mu}^{\rm HVP}$  
by $2.1\,\sigma$.  We show that all these discrepancies could be accounted for by
a new boson produced resonantly around  the  KLOE  centre  of  mass  energy and decaying promptly  
yielding $e^+e^-$ and $\mu^+\mu^-$ pairs in the final states. 
This  gives rise to three different effects:
(i) the additional $e^+e^-$ events will affect 
the KLOE luminosity determination based on 
measurements of the Bhabha cross section, and in turn the inferred value of $\sigma_{\rm had}$;
(ii) the additional $\mu^+\mu^-$ events will affect the determination of 
$\sigma_{\rm had}$ via the (luminosity independent) measurement of the ratio 
of $\pi^+\pi^-\gamma$ versus  $\mu^+\mu^-\gamma$ events;
(iii) loops involving the new boson would contribute directly to the prediction for $a_\mu$.
We discuss in detail this possibility, and we present a simple model that can reconcile the  KLOE and 
BaBar results for $\sigma_{\rm had}$, the data-driven and the lattice determinations of $a_\mu^{\rm HVP}$, 
the predicted and measured values of $a_\mu$,  
while complying with all phenomenological constraints.}

\begin{document} 

\maketitle

\section{Introduction}

The recent experimental result for the muon anomalous magnetic moment $a_\mu$  from the FNAL Muon g-2 experiment~\cite{Abi:2021gix} has confirmed the old BNL measurement~\cite{Bennett:2006fi},  
adding significance  to the long standing discrepancy 
with the standard model (SM) prediction which is now raised to $4.2\,\sigma$.   
Currently, the world  average for this discrepancy is 
\begin{align}
\label{eq:Deltamu}
   \damtwo \equiv a^{\rm exp}_\mu  - a^{\rm SM}_\mu = (2.51 \pm 0.59 ) \cdot 10^{-9} \ ,
\end{align}
where $a^{\rm exp}_\mu$ is the combined result from Refs.~\cite{Bennett:2006fi,Abi:2021gix}, and   
the  SM estimate is the one recommended by the {\it Muon g-2 Theory Initiative}~\cite{Aoyama:2020ynm}
which is mainly based on the estimates in  Refs.~\cite{Aoyama:2012wk,Aoyama:2019ryr,Czarnecki:2002nt,Gnendiger:2013pva,Davier:2017zfy,Keshavarzi:2018mgv,Colangelo:2018mtw,Hoferichter:2019mqg,Davier:2019can,Keshavarzi:2019abf,Kurz:2014wya,Melnikov:2003xd,Masjuan:2017tvw,Colangelo:2017fiz,Hoferichter:2018kwz,Gerardin:2019vio,Bijnens:2019ghy,Colangelo:2019uex,Blum:2019ugy,Colangelo:2014qya}. 
This estimate relies on a data-driven approach that makes use of  experimental measurements of the 
$\shad \equiv \sigma(e^+ e^- \to$ hadrons) cross section to determine the hadronic vacuum polarization 
contribution $\amhvp$. 
This is the most uncertain input  in the prediction for $a_\mu$ and, due to  its 
non-perturbative nature, improving in precision is a  difficult task.
Apart for the uncertainty,  one can also wonder to which level the adopted value can be  considered 
reliable, since 
determinations of $\amhvp$ using data from different experiments exhibit a certain 
disagreement. 
In particular,  KLOE~\cite{Anastasi:2017eio} and BaBar~\cite{Aubert:2009ad} disagree at the level of 3$\,\sigma$, especially in the $\pi^+\pi^-$ channel that accounts for more than 70\% of the value of $\amhvp$, and while  BaBar data  
favour smaller values of  $ \damtwo$,   KLOE data pull  to increase the 
discrepancy.\footnote{Due to relatively larger errors, there is instead agreement within 1.5$\,\sigma$ between  KLOE
and CMD-2~\cite{Akhmetshin:2003zn,Akhmetshin:2006wh,Akhmetshin:2006bx}, 
SND~\cite{Achasov:2006vp}, BES-III~\cite{BESIII:2015equ}.}

The HVP contribution  can also be determined from first principles by means of lattice QCD techniques. 
 However, until recently, the uncertainties in lattice results were too large to 
allow for useful comparisons with the data-driven determination.  
A first lattice QCD computation of   $\amhvp$ with subpercent precision 
was  recently accomplished by the BMW collaboration~\cite{Borsanyi:2020mff}
 $\amhvp=707.5(5.5) \times 10^{-10}$. The result differs from the world average 
 obtained from the data-driven dispersive approach by $2.1\,\sigma$~\cite{Keshavarzi:2021eqa}  
and, in particular, it would yield a theoretical prediction for $a_\mu$  
only $1.3\,\sigma$  below the measurement.\footnote{Other lattice determinations 
also tend to give larger  $\amhvp$  central values although with  considerably larger errors, see e.g the review~\cite{Gerardin:2020gpp}.
} 

We can thus conclude that the muon $g-2$ anomaly~\eqn{eq:Deltamu} 
is accompanied by other discrepancies that, although of lesser significance, 
can shed some suspicion on  interpreting   the $a_\mu$ anomaly as a 
reliable hint of new physics (NP).   
In this respect, 
new high statistics measurements of $\shad$, and in particular in 
the $\pi^+\pi^-$ channel, that might be  soon provided by the CMD-3 
collaboration~\cite{Ryzhenenkov:2020vrk},  as well as new  high precision lattice 
evaluations, which might confirm or correct the BMW result~\cite{Borsanyi:2020mff}, 
will be of crucial importance not only to strengthen or resize the evidences for a $(g-2)_\mu$ anomaly, 
but also to asses the status of the related additional  discrepancies.  
In the meantime, 
it is worthwhile wondering if  the discordant determinations of  $\amhvp$
(KLOE vs. BaBar and $\shad$ vs. lattice) 
could be explained,  jointly    with $\Delta a_\mu$,  within  a single NP scenario.
In this paper we explore this possibility. 

We suggest that the origin of the KLOE-BaBar 
discrepancy could be due to  a NP contribution 
from a hypothetical new resonance $V$ lying close to the $\phi$ mass. This resonance decays semi-visibly  into charged leptons plus additional dark sector 
particles, collectively denoted as $X$, thus producing a certain number 
of $e^+e^-$ and $\mu^+\mu^-$ events 
with a  smooth continuous spectrum. Electron-positron pairs would contribute 
to Bhabha scattering events,  which are used by the experimental collaborations to determine 
the beam luminosity by  comparing   the measured number of $e^+e^-$ events  with the QED prediction. 
Clearly, in those cases in which $\shad$ is determined directly from the number of hadronic 
events~\cite{KLOE:2008fmq}, not accounting for such a NP contribution to Bhabha scattering would 
result in overestimating the luminosity and hence underestimating  
$\shad$. 

On the other hand, events from $e^+e^-\to V \to \mu^+\mu^- X$ would be 
interpreted as $e^+e^- \to \gamma^* \gamma_{ISR}  \to \mu^+\mu^- \gamma_{ISR} $ events in the 
analyses in which the  Initial State Radiation (ISR) photon goes undetected \cite{KLOE:2012anl,BaBar:2012bdw}, and would be counted as 
originating  from processes involving $\gamma^*$ virtual photon exchange,  resulting in an overestimation 
 of the $\sigma_{\mu\mu\gamma}$ cross section from pure QED. Similarly,  events from $e^+e^-\to \gamma V \to \gamma \mu^+\mu^- X$ can also mimic pure QED ISR events in which the  photon is  detected~\cite{BESIII:2015equ,BaBar:2012bdw},  when the missing energy 
 cannot be precisely reconstructed due to experimental limitations.
This would affect the determination of $\shad$ based on the ratio between the  number of $\pi^+\pi^-(\gamma)$ and $\mu^+\mu^- (\gamma)$ events~\cite{KLOE:2012anl}, since 
$    \shad \propto \sigma_{\pi\pi\gamma}/\sigma^{\gamma^*}_{\mu\mu\gamma} > N_{\pi\pi\gamma}/N_{\mu\mu\gamma} 
$, where $\sigma^{\gamma^*}_{\mu\mu\gamma}$ denotes the pure QED process 
that should be used to determine the photon HVP contribution (see \eqn{eq:mumuISR} below).
If these (hypothetical) NP effects are 
instead included, the KLOE vs. BaBar discrepancy can be solved, and  the disagreement between the data-driven 
and the lattice  determination of $\amhvp$ can also be  explained. 

As a concrete example, we discuss a simple model originally 
proposed as a realisation of the ``inelastic dark matter"  mechanism~\cite{Izaguirre:2015zva,Izaguirre:2017bqb,Jordan:2018gcd,Berlin:2018jbm},   
in which the contribution to the Bhabha process $e^+e^- \to e^+ e^-$ 
of a new vector boson $V$ of mass $m_V\sim 1\,$GeV can result in  
an overestimation of the luminosity at KLOE, with negligible effects on the
luminosity determination in  other experiments. Furthermore, the contribution of NP-related 
muonic events  
$e^+e^- \to  (\gamma_{ISR}) V  \to (\gamma_{ISR})  \mu^+\mu^- X $ 
will affect  
the analyses that rely on the ratio $N_{\pi\pi\gamma}/N_{\mu\mu\gamma}$ to determine $\shad$.
In this model,  the values of the  parameters  
required to  account for the  KLOE vs. BaBar discrepancy  and for reconciling  
 the data-driven vs. lattice determination of  $\amhvp$
do not  conflict with other experimental constraints (mono-photon search in BaBar,  $\phi$-related observables, etc\dots). 
The strongest constraints come from 
the KLOE measurement of $\shad$ carried out 
at a center of mass (CoM) energy slightly below the $\phi$ resonance (KLOE10)~\cite{KLOE:2010qei},  
and from the measurement of the 
 forward-backward asymmetry in  $e^+ e^-$ final states~\cite{KLOE:2004uzx}. Yet, these constraints are not sufficiently strong to exclude neither our model, nor the  general NP mechanism underlying it.
While the indirect effects of the new $V$ boson outlined above are still 
 unable to fully account for the $\Delta a_\mu$ discrepancy, 
due to the coupling between $V$ and the muons there is also  a 
direct contribution to $a_\mu$ from loops involving $V$.
This direct effect suffices to bring the theoretical estimate of the muon anomalous magnetic moment in  agreement with the experimental measurement.
Thus, a single NP input is able to solve consistently 
all the $a_\mu$-related anomalies at once. 
 
The paper is structured as follows. In section~\ref{sec:hvp} we first review the main ingredients of the data-driven calculation of $\amhvp$, then we illustrate the mechanism through which a shift in the luminosity determination can affect the KLOE result for $\shad$, and finally we discuss the modification of $\shad$ from NP contributions to $e^+ e^-\to \mu^+ \mu^-\gamma$. In section~\ref{sec:modelandresults} we put forth an explicit realisation based on an inelastic dark matter model,  we study quantitatively  the different effects and we  present all the details 
of   the numerical analysis. Section~\ref{sec:constraints} is devoted to 
an analysis of the existing phenomenological constraints on our NP model. 
In Section~\ref{sec:allinall} we discuss the solution 
to the $a_\mu$-related anomalies, and finally 
in Section~\ref{sec:conclusions} we resume the main results and we draw the conclusions.

\section{The muon magnetic moment and the hadronic cross-section}
\label{sec:hvp}

\subsection{Data-driven calculation of $\amhvp$. }
At leading order, the  contribution to the muon magnetic moment arising from the hadronic 
vacuum polarisation can be derived from the data
on the  hadronic cross-section $\shad =\sigma( e^+ e^- \to \rm hadrons\, (\gamma))$
by means of the optical theorem~\cite{Brodsky:1967sr,Lautrup:1969fr}
\begin{align}
\label{eq:amhvp}
 \amhvplo =   \frac{1}{4\pi^3} 
  \int^\infty_{4 m^2_\pi}
 ds \, K(s) \shad(s) \,.
\end{align}
Here, $\shad$ 
is the bare $e^+e^-\to \gamma^* \to\mathrm{hadrons}\, (\gamma)$ cross-section, obtained by removing 
the infinite string of hadronic vacuum polarisation insertions in the photon propagator 
(that leads to the running of $\alpha_{\rm QED}$), 
and the kernel function reads
\be
    K(x) = \frac{x^2}{2}(2-x^2) + \frac{1+x}{1-x}x^2\log x  + \frac{(1+x^2)(1+x)^2}{x^2} \left( \log(1+x)-x +\frac{x^2}{2}\right)\,,
\ee
with $x=(1-\beta_\mu)/(1+\beta_\mu)$
and $\beta_\mu=\sqrt{1-4m_\mu^2/s}$. 
On the basis of the analyses in
Refs.~\cite{Davier:2017zfy,Keshavarzi:2018mgv,Colangelo:2018mtw,Hoferichter:2019mqg,Davier:2019can,Keshavarzi:2019abf}, 
the   {\it Muon g-2 Theory Initiative}~\cite{Aoyama:2020ynm} has recommended the numerical value  
\begin{align}
 \amhvplo =   693.1 \pm 4.0 \cdot 10^{-10} \,, 
\end{align}
which directly depends  on the accuracy of the hadronic cross-section  experimental determination.
Indeed, the HVP contribution to the muon anomalous magnetic moment is a widely studied subject
(recent works include Refs.~\cite{Davier:2019can,Keshavarzi:2019abf,Colangelo:2018mtw,
Jegerlehner:2017gek,Benayoun:2019zwh,Ananthanarayan:2018nyx}).
In general, various data-sets coming from the same or different experiments and exploiting several final states must be combined. The most important channel is the $\pi^+\pi^-$ channel which accounts for more than $70\%$ of $\amhvplo$. Many experimental measurements have been performed, leading to the embarrassing situation in which  the two most precise measurements disagree at the level of $\sim3\sigma$. Indeed, the values of $\amhvplo$ reported by KLOE 
as the average of their three different analyses~\cite{Anastasi:2017eio} and BABAR \cite{Aubert:2009ad,BaBar:2012bdw}
show a worrisome difference, while other experiments (CMD-2, BESIII, SND \cite{Akhmetshin:2003zn,Akhmetshin:2006wh,Akhmetshin:2006bx,BESIII:2015equ,Achasov:2006vp})
have reported results that lie in between these two, albeit with larger uncertainties. Note that 
any effect that could lift the  overall value of $\shad$ given by KLOE, 
besides reducing the discrepancy with BaBar,  at the same time would also
reduce  $\damtwo$.\footnote{
$\amhvp$ can be also   estimated from  hadronic $\tau$ decays data~\cite{Davier:2010nc,Davier:2010fmf,Davier:2013sfa,Jegerlehner:2017gek,Bruno:2018ono}. The resulting value $(703.0 \pm 4.4)  \times 10^{-10}$~\cite{Davier:2013sfa} is significantly larger than the result  from  $e^+ e^- \to \rm hadrons $,  and in agreement with the lattice estimate.
}

\subsection{Modifying the hadronic cross-section: the luminosity determination}
\label{sec:mod_lumi}

Some of the early experimental results for the 
hadronic cross-section (and in particular the two first analysis from the KLOE collaboration~\cite{KLOE:2008fmq,KLOE:2010qei}, denoted respectively as KLOE08 and KLOE10) depend on the specific  luminosity $\Lee$ of the colliding 
beams: $\shad \propto N_{\rm had}/\Lee $, with 
$ N_{\rm had}$ the number of hadronic events. 
The luminosity  is estimated by comparing high statistics measurements of 
 $e^+ e^- \to e^+ e^-$   events with the SM  prediction for Bhabha scattering:
\be
\Lee^{\rm SM} =\frac{N_{\mathrm{Bha}}}{\sigma_{\mathrm{eff}}^{\mathrm{SM}}},
\ee
where $N_{\mathrm{Bha}}$ is the total number of (background subtracted) Bhabha events. 
For each experiment, the effective SM Bhabha cross section $\sigma_{\mathrm{eff}}^{\mathrm{SM}}$ is evaluated by inserting into detector  
simulations the results of  high precision Bhabha event generators~\cite{Jadach:1996gu,Arbuzov:2005pt,Balossini:2006sd}. 
The presence of  NP contributions to    $e^+ e^- \to e^+ e^-$ unaccounted for by 
$\sigma_{\mathrm{eff}}^{\mathrm{SM}}$
 would then yield an incorrect estimate of the beam luminosity. 
The true luminosity $\Lee$ would  be related to the inferred one as 
\be
\label{eq:lum}
\mathcal{L}_{e^+e^-} = 
\Lee^{\rm SM}  \frac{\sigma_{\mathrm{eff}}^{\mathrm{SM}}}{\sigma_{\mathrm{eff}}} \,,
\ee
where $\sigma_{\mathrm{eff}}$ is the full Bhabha cross-section including 
the NP contribution, that we parametrize in terms of a correction to the QED Bhabha 
  cross section as  
\be
\label{eq:deltaR}
\sigma_{\mathrm{eff}} = \sigma_{\mathrm{eff}}^{\mathrm{SM}} (1+\delta_R)\,.
\ee
Note that a possible difference in the efficiencies for revealing 
SM and NP-related events is also absorbed in  $\delta_R$.
As a consequence, the true luminosity would be smaller (cf. \eqn{eq:lum}),  the 
 true hadronic cross section would become larger   $\shad \to \shad(1+\delta_R)$, 
and thus  the inferred value  of the HVP would be increased as:
\be
\label{eq:deltaamu}
\amhvplo \to \; \amhvplo\,  (1+\delta_R)\,.
\ee
It is understood that the correction $\delta_R$  induced by the shift in the luminosity 
depends on the details of the experimental setup, in such a way that while
it can be quite important in some experiment, it  could be completely negligible
in others. 
Needless to say,   lattice calculations of  $\amhvp$   are not affected by 
this type of indirect effects  on   $\amhvplo$  that stem 
from  modifications of the estimated luminosity.

\subsection{Modifying the hadronic cross-section: the $\sigma(\mu\mu\gamma)$ method}
\label{sec:mod_mumugamma}

Several experimental measurements, including KLOE12~\cite{KLOE:2012anl}, BaBar~\cite{BaBar:2012bdw} and BESIII~\cite{BESIII:2015equ} do not rely on a determination of the luminosity  from 
Bhabha scattering to measure $\shad$, and use instead a luminosity independent method. 
The method exploits the following relation
\begin{equation}
    \label{eq:mumuISR}
    \sigma^0_{\pi^+\pi^-} = \frac{N_{\pi^+\pi^-\gamma_{ISR}}}{N_{\mu^+\mu^-\gamma_{ISR}}}\; \sigma^0_{\mu^+\mu^-}\,,
\end{equation}
that involves the ratio between the number of $\pi^+\pi^-$ and $\mu^+\mu^-$ events 
with an ISR photon, and the $\sigma^0_{\mu^+\mu^-}\equiv \sigma(e^+e^- \to \mu^+\mu^-)$ cross section 
computed in QED.
The KLOE12 measurement is peculiar in that it does not detect the photon, since 
the angular cuts that are used to enhance  ISR  over final state radiation (FSR) events imply that 
the photon gets lost in the beam pipe. The CoM energy of the collision is then reconstructed 
from  the charged particles invariant mass.

If we have an excess of NP-related  $\mu\mu X$ 
events
(where $X$ represents undetected particles)  mimicking  $\mu \mu \gamma$  
final states and contributing to the 
denominator 
in \eqn{eq:mumuISR},  these events need  to be subtracted out in order that 
the derived value of $\sigma^0_{\pi^+\pi^-}$ could be correctly related to the photon HVP.
The  hadronic cross-section induced by virtual photon exchange is then shifted as 
\begin{align}
\label{eq:delmudelpi}
    \sigma^{0}_{\pi^+\pi^-} \quad \longrightarrow \quad  
      \sigma^{0\,(\gamma^*)}_{\pi^+\pi^-} \; \simeq\; 
    \sigma^0_{\pi^+\pi^-} \left[1 + \delta_\mu (s')\right] \,, 
\end{align}
where $s'$ is the invariant mass squared of the charged particles, and we have defined:
\begin{align}
\label{eq:dmu}
\delta_\mu (s')~\equiv~ \frac{\sigma^{NP}_{\mu\mu X}  (s') }{ \sigma_{\mu\mu}  (s')} \frac{\epsilon^{NP}}{\epsilon^{SM}}\ ,
\end{align}
where $\epsilon^{NP}$ ($\epsilon^{SM}$) is the efficiency of the selection cuts for the NP (SM) events.
The cross section $\sigma^{NP}_{\mu\mu 
X}$ represents the NP contribution yielding events for which the final state may contain 
additional (invisible) states, and that are counted in the experimental analysis  as long as they pass the  kinematic cuts.

The shift in the hadronic cross section depends on $s'$ and  has to be integrated along with the kernel function $K(s')$ in order to obtain the final shift to $\amhvplo$.
A similar correction will also affect every measurement which uses the di-muon final state as a luminosity measurement. Finally, a flavour-universal new physics effect capable of modifying the large angle  $e^+ e^- \to e^+ e^-$ cross-section at the level of a few percent, can in principle lead to an effect of similar size for 
$\gamma \mu \mu  X$ events, where an ISR photon is emitted, up to  differences 
related to the muon mass and to different experimental cuts. Therefore
we can expect the correction $\delta_R$ of the previous section and 
the correction $\delta_\mu$ discussed in this section to be of a similar size. 

A remark is in order regarding the possibility that the NP will also 
produce an excess of hadronic events. Since these events will not be related 
to the exchange of a virtual $\gamma^*$, they must be subtracted from the 
hadronic data sample in order to  reconstruct 
the $  \sigma^{0\,(\gamma^*)}_{\pi^+\pi^-} $ cross section from which the 
photon HVP can then be derived. 
Let us  define in complete analogy 
with \eqn{eq:dmu} a correction 
\begin{align}
\label{eq:dpi}
\delta_\pi (s')~\equiv~ \frac{\sigma^{NP}_{\pi\pi X}  (s') }{ \sigma_{\pi\pi}  (s')} \frac{\epsilon^{NP}}{\epsilon^{SM}}\,.
\end{align}
The effect of the $\pi\pi$ events of NP origin can then be  subtracted out  
simply by replacing in \eqn{eq:deltaR}  (and in \eqn{eq:deltaamu}) 
$\delta_R\to \delta_R - \delta_\pi (s')$, and in \eqn{eq:delmudelpi}
$\delta_\mu(s')\to \delta_\mu(s') - \delta_\pi (s')$.
It is clear that the relative strength characterising the NP 
couplings to the leptons and to the hadrons is a model dependent issue, 
so that the  size of the hadronic shift $\delta_\pi$ relative to the leptonic ones
$\delta_R, \delta_\mu$
cannot be assessed in a general way. 
However,  in our signal region, corresponding to the $\rho / \omega$ resonance, the SM cross-section for hadronic final states is larger than 
the  one for muons by about an order of magnitude. Therefore, in models 
where the NP controbution to the  leptonic and hadronic  channels are of similar size,
we have 
$\delta_R \sim \delta_\mu (s')\gg \delta_\pi (s')$. This will be the case 
in the example that will be detailed below, where  
in first approximation the hadronic shift can be neglected, and we can simply set $\delta_\pi(s') \approx 0$.

\subsection{Overall effects of new physics on $\amhvp$}
\label{sec:globalfit}

Several experiments have provided measurements of the hadronic cross section. Combining 
these measurements is an intricate process which requires the use of dedicated  codes, which 
are not  publicly available~\cite{Keshavarzi:2019abf,Davier:2019can}. 
In order to get a handle of the soundness of the numerical procedure that we have adopted, 
we have relied on the available results for the various experiments   reported  
in Refs.~\cite{Keshavarzi:2019abf,Davier:2019can},
focusing on the range $\sqrt{s'} \in [0.6, 0.9] \, \rm GeV$ for the reduced energy:  
\begin{align}
\label{eq:expresults}
    \amhvp ( \sqrt{s'} \in [0.6, 0.9] \, \rm \, GeV ) = 
    \begin{cases} 
    ( 366.9 \pm 2.1) \cdot 10^{-10} \rm \quad (KLOE) \\
    ( 376.8 \pm2.7 )\cdot 10^{-10} \rm \quad (BABAR) \\
    (372.5 \pm3 )\cdot 10^{-10} \rm \quad (CMD-2) \\
    (368.3 \pm4.2 )\cdot 10^{-10} \rm \quad (BESIII) \\
    (371.8 \pm5 )\cdot 10^{-10}\rm \quad (SND) \\
    (377.0 \pm6.3 )\cdot 10^{-10} \rm \quad (CLEO) \, .
    \end{cases}
\end{align}
When combined with a simple $\chi^2$, we obtain a global fit of $\amhvp ( \sqrt{s'} \in [0.6, 0.9] \, \rm GeV ) = 371.1$ that  matches the results of both the combined fit presented  in Ref.~\cite{Davier:2019can,Keshavarzi:2019abf} (see also table 6 in Ref.~\cite{Aoyama:2020ynm}),  thus corroborating the reliability  
of our method.

Feeding the indirect effects of NP discussed in Sections \ref{sec:mod_lumi} and \ref{sec:mod_mumugamma}
into the results listed in \eqn{eq:expresults} is far from being a straightforward step. 
This is 
because different measurements are affected by the NP in different ways, while others are not affected at all. 
In particular, the KLOE result in \eqn{eq:expresults} corresponds to an average of three different analyses,  
labelled as KLOE08~\cite{KLOE:2008fmq}, KLOE10~\cite{KLOE:2010qei} and KLOE12~\cite{KLOE:2012anl}. 
A NP-related shift in the luminosity determination (see Section~\ref{sec:mod_lumi}) 
can affect KLOE08 and KLOE10, but not KLOE12. This last analysis  
can instead be affected by  additional NP-related $\mu^+\mu^-$ events (see Section \ref{sec:mod_mumugamma}). 
The rest of this section is thus devoted to unfold the details of the indirect NP effects on the three   KLOE analyses.

 The 
 hadronic cross section $\sigma^0_{\pi\pi}(s')$, whose determination was the aim of the 
  KLOE analyses, is related to  the radiative cross section that includes the emission of an ISR photon 
 via the following relation: 
    \be
    s\frac{d \sigma(\pi^+\pi^-\gamma)}{d s'}=\sigma^0_{\pi\pi}(s') H(s',s),
    \label{eq:KLOEshad}
    \ee
    where $H$ is the radiator function that accounts for the ISR  and $s'=M^2_{\pi\pi}$ is the di-pion invariant mass.
    The different strategies followed by the KLOE collaboration in their three different analysis are resumed as follows: 
\begin{itemize}
    \item 
        KLOE08 data-set consists on 60 measurements in the range $0.35 < s'/\mathrm{GeV}^2 < 0.85$ collected 
        with the $e^+e^-$ beam energies tuned to  $\sqrt{s}=1.0194\,$GeV, that is at the $\phi$-meson pole. 
    Their determination of $\sigma^0_{\pi\pi}$ relies directly on Eq. (\ref{eq:KLOEshad}), 
    where the radiative cross section in the LH side is inferred from the number of $\pi^+\pi^-\gamma$ events 
    that pass the cuts, and relies on the knowledge of the luminosity that, as discussed in Section \ref{sec:mod_lumi}, 
    is measured via Bhabha scattering.

     \item  
          KLOE10 data-set consists of 75 measurements of the radiative cross section 
          in the range $0.1 < s'/\mathrm{GeV}^2 < 0.85$, collected 
          with the $e^+e^-$ beam energies tuned to  $\sqrt{s}=1\,$GeV, that is 
          about $4.5\times\Gamma_\phi$ below the $\phi$ pole. Also in this case  the inferred  hadronic cross section   
          depends on the measurement of the luminosity. However, given the $\sim 20\,$MeV difference in the CoM energy  
          with respect to KLOE08, the two measurements could be simultaneously affected by a luminosity 
          redetermination only if  the width of the new resonance is  
           several times larger than $\Gamma_\phi$. The model discussed in the next section yields instead a width of order  $\Gamma_\phi$. Then it turns out  that the most convenient choice is that of shifting KLOE08 results
           by locating the new resonance close to $m_\phi$,  while leaving the KLOE10 measurement unaffected by the NP.

      \item 
     KLOE12 relies instead on \eqn{eq:mumuISR} where  data correspond to the ratio 
     between the number of $\pi^+\pi^-\gamma$ and $\mu^+\mu^-\gamma$ events 
     (where the photon goes undetected) measured in 30 different bins in the 
     interval  $0.35 < s'/\mathrm{GeV}^2< 0.95$, 
     where $s'$ corresponds to the $\mu\mu$ and $\pi\pi$ invariant  mass.  
      The advantage of this method is that  
      the dependence on the luminosity cancels in the ratio, so that  
      KLOE12  is not affected by possible changes in the luminosity determination. However, as described in Section \ref{sec:mod_mumugamma}, the measurement would be directly affected by additional $\mu^+\mu^-$ events of NP origin accompanied by  undetected  $X$  particle(s). 
\end{itemize}

In the invariant mass squared range in which the three KLOE analyses overlap, the numerical results 
for the individual contribution to the muon magnetic anomaly are~\cite{Anastasi:2017eio}:
\begin{align} 
\label{eq:KLOEall}
    \amhvp ( s' \in [0.35, 0.85] \, \rm \, GeV^2 ) = \begin{cases} 
    ( 378.9 \pm 3.2) \cdot 10^{-10} \rm \quad (KLOE08) \\
    ( 376.0 \pm 3.4 )\cdot 10^{-10} \rm \quad (KLOE10) \\
    (377.4 \pm 2.6 )\cdot 10^{-10} \rm \quad (KLOE12)  \, ,
    \end{cases}
\end{align}
We use these determinations and, after correcting them for the effects of 
$\delta_R$ (luminosity shift) and $\delta_\mu$ ($\mu\mu$  events of NP origin)
 we combine them into a single KLOE result.\footnote{As discussed in~\cite{Anastasi:2017eio} there 
 are large correlations among the three different KLOE determinations of   $\amhvp$. 
Most sources of systematic uncertainties, which represent the leading contribution to the 
total error, are fully correlated between all energy bins. 
There is  also a full statistical correlation in the two-pion data that are shared 
between KLOE08 and KLOE12~\cite{Anastasi:2017eio}. 
 We defer to future work the complete analysis of the KLOE datasets corresponding to a total of 
 $195$ data bins along with their correlation matrices.} 

While we have so far not specified an explicit NP mechanism responsible for the indirect 
corrections, we anticipate that we have in mind the excitation of a new resonance which  promptly
decays, yielding the required excess of $e^+e^-$ and $\mu^+\mu^-$ pairs possibly accompanied by other
undetected particles.
As a consequence, a measurement performed at a CoM energy below the NP resonance would remain unaffected 
by these corrections. However,  experiments that exploit the ISR method to scan over the reduced 
energy $\sqrt{s'}$  will generally be affected whenever the CoM energy $\sqrt{s}$ is above the resonance, even if    
not  particularly close to the pole. This is because 
ISR photon emission can downgrade the effective $e^+e^-$ collision energy to the right value to excite the resonance. 
Hence, for example,  the BESIII measurement ($\sqrt{s} \simeq 3.8\,$GeV,  
 $ \sqrt{s'} \in [0.6,\,0.9] \,$GeV) would  be affected by a new resonance 
sitting close to the $\phi$ mass, and in our study we have taken this effect into account.  
BaBar data-taking was carried out at $\sqrt{s} \simeq 10.6\,$GeV, and $\amhvplo$ was  calculated using 
the measured $\pi^+\pi^-$  cross section from threshold to $1.8\,$GeV, and in this case the high energy bins would clearly 
not get affected by a resonance at the $\phi$ mass. Conservatively, 
we have not included in our main results a possible correction to the BaBar measurements, 
even for the subset of data restricted to the $\sqrt{s'}$ interval in \eqn{eq:expresults}.  
However,  to get an handle on the possible consequence of shifting also the BaBar measurement,  
we will show the effects of assuming a correction half the size  
the one for BESIII.

Finally, we should also stress that our analysis has been forcefully restricted  to the  
invariant mass squared interval $  0.35\leq s'/\mathrm{GeV}^2 \leq 0.85$  for which we could  
 carry out our procedure in a consistent way. 
However, since the contribution of the $e^+e^-\to \pi^+\pi^-$ channel to $\amhvp$ over the complete $ \sqrt{s'}$ 
range is about $35\%$ larger, we expect that a full fit using the dedicated 
(but non-public)  codes from~\cite{Keshavarzi:2019abf,Davier:2019can} could lead to a sizeably 
larger shift in $\amhvplo$.  This means that the results we are presenting here 
can be taken as a conservative estimate of the possible NP effects on the $ \amhvp$ determination.

\begin{figure}[tb]
\centering
\includegraphics[width=0.85\textwidth]{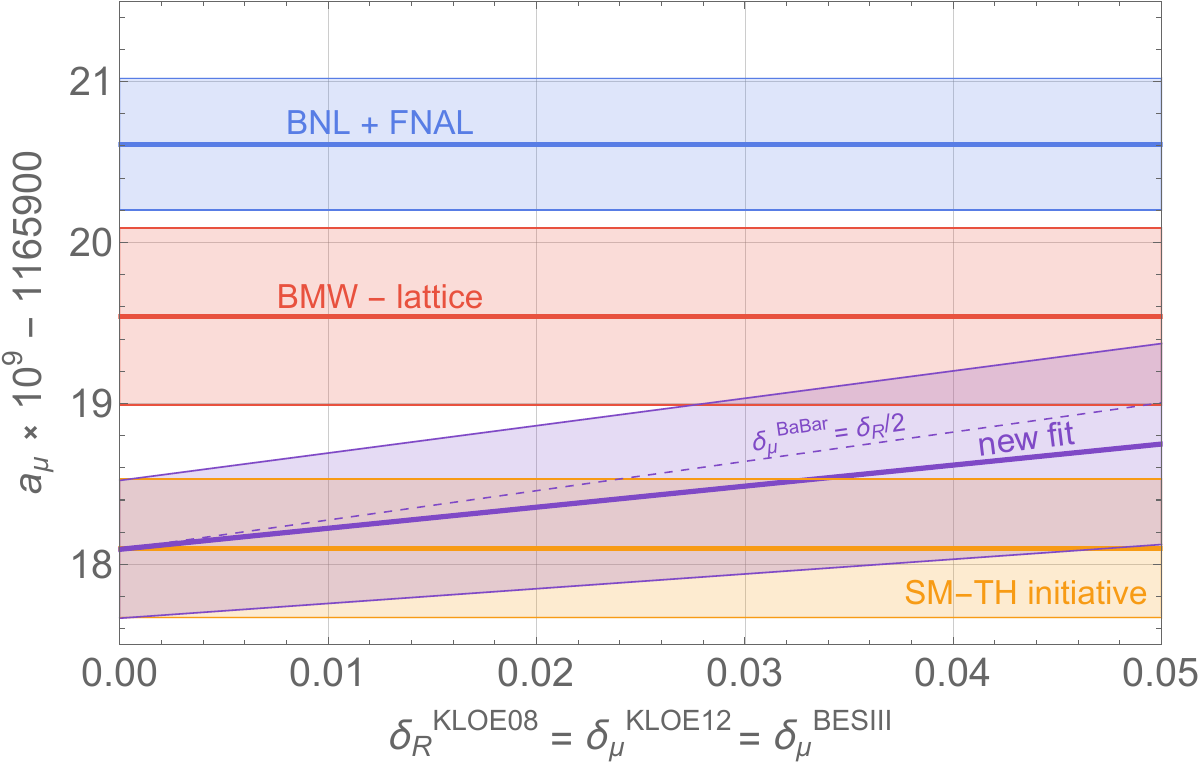}
\caption{Theoretical prediction for $a_\mu$ based on a  
data driven global fit to $\amhvplo$ obtained by including a NP correction  
affecting in an equal way $e^+ e^-$ and $\mu^+ \mu^- \gamma$ final states (oblique violet band). 
The thick purple line assumes that the NP does not affect the BaBar result, while the 
dotted purple line assumes for BaBar an effect  half the size than the one in KLOE and BESIII. 
The blue band corresponds to the combined BNL and FNAL  experimental results, the red band to 
the  prediction obtained with the BMW lattice estimate of $\amhvplo$~\cite{Borsanyi:2020mff}, and 
the orange band to the one obtained from $\sigma_{\rm had}$ in the absence of NP corrections. 
The  width of the bands represent $1\,\sigma$ uncertainties.} 
\label{fig:newFit}
\end{figure}

We illustrate the results of our fit  in Fig.~\ref{fig:newFit}. The blue band corresponds to the combined BNL and FNAL experimental results, the red band to the prediction obtained with the BMW lattice estimate of $\amhvplo$~\cite{Borsanyi:2020mff}, and the orange band to the one obtained from $\shad$ without modifications of the KLOE and BESIII results. 
The excellent  agreement  between our $\chi^2$ fit and the full results of Refs.~\cite{Keshavarzi:2019abf,Davier:2019can} 
is put in evidence  by the precise overlap between the violet and the orange bands occurring at the left boundary of Fig.~\ref{fig:newFit}, where the KLOE and BESIII results are taken at face value and not modified.
We next increase the KLOE08, KLOE12 and BESIII value for $\sigma_{\rm had}$ assuming this is due to some unspecified NP contribution that affects the luminosity measurement (KLOE08) and the detected number of $\mu\mu\gamma$ events (KLOE12 and BESIII), while it remains negligible for all the other measurements. The purple band shows how a  2-3 $\%$ 
shift of equal size in KLOE08, KLOE12 and BESIII would yield  a shift of $\sim 0.5 \times  10^{-9}$ in $\amhvplo$, 
and would drive the dispersive determination of $\amhvplo$ in agreement 
at the $\sim 1.5\sigma$ level with the BMW-lattice result. As will be discussed in the following sections,  
a moderate shift of this size 
would also maintain the different KLOE analyses in overall satisfactory agreement among them.

On the other hand, in order to reduce  to the $\sim 1\,\sigma$ level the   tension with the  
combined BNL and FNAL experimental results, $\amhvplo$ 
must be increased by an appreciably larger factor $\sim 1.6\cdot 10^{-9}$.
However, this would bring  the KLOE08 and KLOE12 results in serious disagreement with KLOE10,
and  KLOE12 data on  $\mu \mu \gamma$ will also get in strong tension with the SM prediction. 
Finally, the dotted purple line assumes a shift for BaBar of half the size the one in BESIII and KLOE.

To summarise, a NP effect acting only indirectly through a luminosity shift (KLOE08) and via the production 
of an excess of  $\mu\mu\gamma$ events (KLOE12 and BESIII) would suffice to reinstate a satisfactory agreement 
between the different experimental determinations of $\shad$, and to reconcile the dispersive data-driven and 
lattice estimates of $\amhvplo$. Yet, it would not 
be able to  fully solve  the discrepancy between the measured value of $a_\mu$ and the theoretical prediction, which
thus still calls for a direct NP contribution to the muon anomalous magnetic moment. 
In the next section  we will present and explicit example in which 
 a new Feebly Interacting Particle (FIPs) produced resonantly at the 
 KLOE08/KLOE12 CoM energy  and decaying mostly ``semi-visibly'' into $e^+ e^-$ and $\mu^+\mu^-$, 
 can produce indirect effects of the required size, 
 while a direct loop contribution of the FIP to the the muon anomalous 
 magnetic moment can eventually solve all the  $a_\mu$-related anomalies.

\section{Model realisation and analysis}
\label{sec:modelandresults}
In section~\ref{sec:model} we describe
a simple model, that fits in the so-called ``inelastic dark matter (iDM) scenario", which is well 
suited to  increase 
the number of Bhabha events in KLOE08 and to provide additional $\mu^+\mu^-$ events in KLOE12, 
while evading at the same time all other experimental constraints. 
In section~\ref{sec:numericslum} we first review the details of the luminosity measurements in the different experiments that 
contributed  to the determination of $\shad$, and then we analyse  quantitatively 
the effects on the  determination of $\amhvplo$ from $\shad$ data that could result from an overestimation  of the KLOE luminosity. 
In section~\ref{sec:numericsmumu} we study the effects of additional NP-related $\mu\mu X$ 
events at KLOE12. 

\subsection{An inelastic dark matter model }
\label{sec:model}

We introduce a ``dark'' Abelian gauge group $U(1)_D$ with gauge coupling $g_{D}$, along with a dark Higgs $S$ with  $U(1)_D$-charge $q_S=+2$, and two Weyl spinors $\eta,\xi$ 
   with  charges  $q_{\eta} =  -  q_{\xi} = - 1 $   that can be combined in a 
  Dirac fermion $\chi = (\eta\; \xi^\dagger)^T$. 
  We furthermore assume that a charge conjugation symmetry extended to the dark sector  
  enforces invariance under the transformations $V \to -V$,  $ \eta \to \xi^\dagger$
 and $\xi^\dagger \to \eta$~\cite{Berlin:2018jbm}.

Under these premises, we can write the following Lagrangian terms: 
\begin{align}
	\mathcal{L}_V &= -\frac{1}{4}F^{\prime\mu\nu}F^{\prime}_{\mu\nu}
	-\frac{g^\prime \varepsilon}{\cos\theta_w} V_\mu \mathcal{J}_Y^\mu\ , \\ 
	\mathcal{L}_S&=(D^\mu S )^\dagger(D_\mu S)+ \mu_S^2 |S|^2 - \frac{\lambda_S}{2} |S|^4 - \frac{\lambda_{SH}}{2} |S|^2  |H|^2 \ , 
	\label{ea:lagmass}
\end{align}
where we have introduced the customary kinetic mixing term $\varepsilon$ corresponding to a small interaction  between the dark photon $V_\mu$ and the hypercharge current 
$ \mathcal{J}_Y^\mu$~\cite{Holdom:1985ag,Fayet:2016nyc}.
The Lagrangian for the dark fermions can be written as 
\begin{align}
	\mathcal{L}^{\rm DM}&=\bar{\chi}\left( i\slashed{D}-m_{\chi}\right)\chi -
	\frac{1}{2} y_{S} S (\eta^2 + {\xi^\dagger}^2 ) + \textrm{ h.c.} 
\end{align}
where the second term describes the coupling 
between the Weyl fermions and the dark Higgs boson $S$.
Note that the  charge conjugation symmetry mentioned above enforces that 
both $\eta$ and $\xi^\dagger$  couple to $S$ with the same Yukawa coupling. 
The dark Higgs boson mass $m_S$ and the dark photon mass $m_V$ read
\begin{align}
\label{eq:mass}
 m_S &= \sqrt{2 \lambda_S}\, v_S \ , \\
m_V &= 2 g_{D}  v_S = \left( \frac{\sqrt{2} g_D}{\sqrt{ \lambda_S}} \right)  m_S\ .
\end{align}
The diagonalisation of the fermion mass matrix is straightforward and leads to two  states 
$\chi_1 = \frac{i}{\sqrt{2}} (\eta -\xi)$ and 
$\chi_2= \frac{1}{\sqrt{2}} (\eta +\xi)$  with masses 

\begin{align}
    m_{\chi_1,\chi_2} = m_{\chi}\mp \frac{1}{\sqrt{2}} y_S v_S  \ .
\end{align}
The phase of $\chi_1$ has been fixed to obtain a positive mass term under the assumption
$ m_{\chi}\gsim y_S v_S /\sqrt{2}$.
In the mass basis, the dark Higgs $S$ couples diagonally to $\chi_{1,2}$ (expressed in term of the Majorana spinors ) via the term:  
\begin{align}
\mathcal{L}_{S\chi} =-  \frac{1}{2} y_S\, S\,  (\chi_2^2 -\chi_1^2) \,,
 \label{eq:Schi}
\end{align}
where the Yukawa coupling can be expressed as 
\be
\label{eq:yukaS}
y_S= \frac{g_D}{\sqrt{2}} \frac{m_{\chi_2}-m_{\chi_1}}{m_V}
\ee
In contrast, the dark photon interacts with the  fermions via an  off-diagonal coupling 
\begin{align}
\mathcal{L}_{V\chi} =-i g_D V_\mu \bar \chi_2 \gamma^\mu \chi_1 \ .
 \label{eq:Vchi}
\end{align}
This construction can satisfy the main  conditions to generate a significant shift in the KLOE luminosity estimate and to provide additional di-muon events,
while escaping detection in other experiments:
\begin{itemize}
    \item the dark photon mass must be very close to the KLOE CoM energy $\sqrt{s} \simeq 1.02 \, \rm GeV$, in order to produce $V$ resonantly;
    \item dark photon decays must contribute non-negligibly to Bhabha scattering events, and therefore they need to  
     include   $e^+ e^-$ pairs with  invariant mass close to $1$ GeV, as well as additional di-muon events;
    \item in order to escape bump searches, the dark photon main decay channel must be multibody and must include some missing energy.
\end{itemize}
We can choose $ m_V \sim 1 \, {\rm GeV} \gtrsim m_{\chi_2} \gg m_{\chi_1} $ by fixing $v_S$ and 
by requiring the approximate relations
\be
    y_S v_S \sim   \sqrt{2} m_{\chi}\,,  \quad 
     \sqrt{2} y_S \lesssim g_D \,.
\ee
When this arrangement of mass values is satisfied, the dark photon decays 
mainly  proceed through the  chain 
$V \; \to\; \chi_1\, \chi_2 \; \to \; \chi_1 \, \chi_1\,  e^+ e^-\; (\mu^+\mu^-)$.
That is,  $V$ decays   with  almost a $100 \%$ branching ratio into $\chi_1 \chi_2$
while direct $V\to e^+e^-\; (\mu^+\mu^-)$ decays are suppressed as $\varepsilon^2$  
and thus subdominant. 
The  $\chi_2 \to \chi_1\, (\textrm{hadrons})$ decay will also be present at 
a similar level than the (semi)leptonic channels (see e.g. Fig. 1 of ~\cite{Batell:2021ooj}). As argued in Sec.~\ref{sec:mod_mumugamma}, 
due to the fact that the SM hadronic cross section is much larger 
than the leptonic ones, the relative NP correction to the hadronic channel 
will be accordingly suppressed. For this reason we will not include it in our simulations.

 The lightest new fermion $\chi_1$, which may play the role of the dark matter particle, is stable and escapes undetected, while the heavier 
fermion $\chi_2$ decays  into $\chi_1\, e^+ e^-\; (\mu^+\mu^-)$, providing the main contribution to the NP-related 
additional electron-positron (di-muon) pairs.
The  $V$ width that is almost saturated by the $V \to \chi_1 \chi_2$ process  will be an important quantity 
 in the rest of this work. In the limit  $\delta m = m_V-(m_{\chi_1}+m_{\chi_2}) \ll  m_V $ we have:
\begin{align}
\label{eq:widthV}
  \Gamma_{V}
  \simeq \frac{2 \alpha_D \sqrt{\delta m} (\delta m + 2m_{\chi_1})^{3/2}}{m_V} \ .
\end{align}
In particular,   in the benchmark points used in the next section,  phase space suppression 
due to $m_{\chi_2}\sim m_V$  leads to a width 
of order  MeV.
Note that since there are charged particles in the decay chain final state, 
the $V$ boson  escapes invisible decays searches. At the same time,  since 
the three body decay $\chi_2 \to \chi_1  e^+ e^-$ is characterised  by a 
continuous and smooth $e^+e^-$ energy spectrum, it also escapes standard ``bump'' searches.

The mass of the dark Higgs boson depends on its quartic coupling $\lambda_S$,  which  is a free parameter in the theory. 
The particular case $ m_S \lsim 2 m_{\chi_1}$, which ensures that the 
$S\to \chi_1\chi_1$ decay channel is closed, represents an interesting 
possibility to render $\chi_1$ a good DM candidate~\cite{Darme:2017glc,Darme:2018jmx,Wojcik:2021xki}.
The argument proceeds in two steps: 
\begin{itemize}

\item[1.] The $t$-channel $p$-wave  $ \chi_1 \chi_1 \leftrightarrow S S $ process is enhanced by a relatively large Yukawa coupling (see \eqn{eq:yukaS}). As a consequence, it keeps 
 $\chi_1$ in equilibrium  with the $S$ population up to a temperature $T_{\mathrm{fo}}$. This temperature can be modified by changing the $S$ mass in the regime $m_S \gtrsim m_{\chi_1}$.
 
  \item[2]  At $T \gtrsim m_S $ thermal equilibrium between the dark sector 
  composed of $S$ and $\chi_1$ and the lightest SM particles is typically maintained via the standard $\chi_2 \chi_1 \to e^+ e^-$ process, with the $\chi_1$ relativistic. 
Moreover, at $T < m_S$, the contribution of $S$  decays and inverse decays   mediated by a triangle loop  involving $V\,V\,e$ maintains $S$ in thermal equilibrium.\footnote{Note that these  loop processes typically dominate over the Higgs-portal induced interactions due to the large kinetic mixing considered in this study (see e.g.~\cite{Darme:2017glc}).}
Consequently,  $S$  continues to be coupled to the thermal bath 
down to temperatures  $T\ll m_S$ and in particular 
throughout  the  $\chi_1 \chi_1 \to S S$  annihilation process.
    
\end{itemize}
Arranging for   freezing-out the  $\chi_1 \chi_1 \to S S$ process around $T_{\mathrm{fo}}\sim m_{\chi_1}/20$  then allows to reproduce  the correct dark matter relic density. This corresponds to the ``forbidden annihilation'' regime of the iDM model which can be realised  when $\frac{M_S}{2}\lesssim  m_{\chi_1} \lesssim M_S$.\footnote{In the case where the $S$ boson has a longer life-time, additional effects such as the dilution of the $\chi_1$ relic density from the entropy injection from $S\to e^+ e^-$ could also lead to the proper relic density, see the recent work~\cite{Asadi:2021bxp}, although additional constraints from BBN then apply (see e.g.~\cite{Darme:2017glc,Fradette:2017sdd,Kawasaki:2017bqm}}

Finally, let us briefly discuss the direct contribution of  $V$ to  the muon $g-2$. 
Any vector  particle with a direct interaction with muons will give rise to a one-loop contribution to $a_\mu$. For a pure vector coupling, this contribution  can be written as:
\begin{equation}
\label{eq:gm2V}
      \damtwo  =  \frac{\alpha_{\rm em} \varepsilon^2}{2 \pi} \,    x_\mu^2 \; \mathcal{F} 
      \left(x_\mu \right)  \,, 
\end{equation}
where  $x_\mu = m_\mu/m_V$, $g_{eV} = \varepsilon e$ with $e$ the electromagnetic coupling constant,  
and the loop function $\mathcal{F}$ 
is given by:
\begin{align}\label{eq:gm2loop}
\mathcal{F}(x) &=\int_0^1 dz\, \frac{2 z^2 \left(1-z\right)}{x^2 z + (1-z)(1-x^2z)}\,.
\end{align}
For $m_V$ in the GeV range, the loop function simplifies and one has approximately
\begin{align}
     \damtwo \sim 2 \cdot 10^{-9} \times \left(\frac{ e \varepsilon }{0.005}\right)^2  \times \left( \frac{1 \, \rm GeV}{\Mv}\right)^2\,. 
     \label{eq:g-2direct}
\end{align}
The direct contribution in Eq.~\eqref{eq:g-2direct}, together with the indirect corrections from the luminosity determination, can reconcile the theoretical predictions for $a_\mu$ with the experimental results.

\subsection{Numerical analysis: luminosity determination}
\label{sec:numericslum}

 We have implemented the iDM model in \fr/UFO~\cite{Christensen:2009jx,Degrande:2011ua,Alloul:2013bka} files, and we have used the \amc\ platform~\cite{Alwall:2014hca} in order to generate 
  $s$- and $t$-channel  $e^+e^-$ events\footnote{We simulate $e^+ e^- \to \chi_1 \chi_2 , \chi_2 \to \chi_1 e^+ e^-$, including the leptonic branching ratios of the $\chi_2$.}      
that would  contribute to the KLOE, BaBar and BESIII  measurements of the  Bhabha 
cross section.\footnote{ The two experiments CMD-2~\cite{Akhmetshin:2001ig,Akhmetshin:2003zn} and SND~\cite{Achasov:2005rg,Achasov:2006vp} carried out measurements in the energy range for $\pi^+\pi^- (\gamma)$ production by scanning directly with the beam energy. For each point,  Bhabha 
    events were used to calibrate the luminosity. The angular cuts applied are respectively
$    \cos \theta \in [-0.83,0.83] \ \text{(SND~\cite{Achasov:2005rg})} $ and 
        $ \cos \theta \in [-0.45,0.45] \ \text{(\textrm{CMD-2}~\cite{Akhmetshin:2003zn})} \,$.
    These experiments differ in two main aspects from the previous ones. First they cannot distinguish 
    easily $\pi$'s from $\mu$'s, so that the final hadronic cross-section can be obtained only after subtracting the $e^+e^- \to  \mu^+ \mu^-$ component by  relying on the  theoretical estimation. Second, each  energy point in the scan has  its own luminosity measurement, so that the effects of NP  would need to be estimated independently for each $\sqrt{s}$. } We have simulated the process $ e^+ e^- \to \chi_1  \chi_1 e^+ e^-$ and we have applyed the relevant cuts directly on the generated final states. 
  For the luminosity measurement, the different experiments exploited various kinematic ranges:
\begin{itemize}
    \item At KLOE, with CoM energy $\sqrt{s} = 1.02 \, \rm GeV$, the following cuts were applied \cite{KLOE:2006itf}: $\cos \theta \in [-0.57,0.57]$, $E_e \in [0.3,0.8] \, \rm GeV$, $p \geq 400$ MeV and a cut on the polar angle acollinearity $\zeta < 9^\circ$.
     KLOE  reported the Bhabha cross section  $\sigma^{\rm vis}_{e^+e^-} = (431\pm 0.3)$ nb. 
    
    \item At BaBar, with asymmetric collisions of $9$ GeV electrons and $3.1$ GeV positrons corresponding to $\sqrt{s} =  10.58 \, \rm GeV$, the following cuts were applied~\cite{Lees:2013rw}: the polar angles in the centre of mass are required to satisfy $|\cos\theta| < 0.7$ for one track and $|\cos\theta| < 0.65$ for the other; the scaled momentum $P_i=2p_i/\sqrt{s}$, where $p_i$ are the momenta of the track $i$ and $\sqrt{s}$ is the CoM energy, were required to satisfy $P_1>0.75$ and $P_2>0.5$ ($i=1$ denotes the track with the higher CoM momentum); the 
    cut on the acollinearity angle was $\zeta < 30^\circ$. 
     The luminosity  was  estimated also by exploiting the $e^+e^- \to \mu^+ \mu^-$ process and,  
    according to the uncertainties reported in Table 3 of~\cite{Lees:2013rw}, the muon process 
    slightly dominates the average.
     Furthermore, BaBar associated a $0.7 \%$ uncertainty on the Bhabha estimates due to ``Data-MC differences''.  After  accounting for the detector efficiency, 
    the reported measurement for the Bhabha cross section is $ \sigma_{e^+e^- \to e^+ e^-}^{\rm vis} = (6.169 \pm 0.041) \, \rm nb \,$
 and  
that for the $e^+e^- \to \mu^+ \mu^-$ process is 
$\sigma_{e^+e^- \to \mu^+ \mu^-}^{\rm vis} = (0.4294 \pm 0.0023) \, \rm nb $. 

    \item The BESIII collaboration also delivered an initial state radiation (ISR) measurement~\cite{BESIII:2015equ}, although with larger uncertainties, at the CoM energy  $\sqrt{s} =  3.773 \, \rm GeV$. This measurement 
    is in better agreement  with KLOE data. The luminosity estimate~\cite{Ablikim:2014gna} relies solely on Bhabha scattering, with the relatively loose cuts $\cos \theta \in [-0.83,0.83]$. They reported a Bhabha cross section of $\sigma^{\rm vis}_{e^+e^-} = (147.96 \pm 0.74)$ nb.
\end{itemize}

Because of the strong suppression of the Bhabha scattering due to the angular cuts and the $\varepsilon^4$ suppression of direct leptonic $V$ decays, the order percent NP contribution required to shift significantly $\amhvp$ cannot be obtained from the off-shell exchange of a $V$ boson.
However, for $s$-channel on-shell $V$ production  the  suppression is partly compensated by 
 the resonant nature of the process.
The direct consequence is that  given a vector boson $V$ of a certain mass, at most one among the above experiments can be affected non-negligibly in  its luminosity determination.

  Since the KLOE measurement is the main responsible of reducing the 
  data driven value of the HVP,  we need to increase $\shad$ around the CoM energy 
  of $\sim 1\,$GeV. Accordingly, we require a dark photon mass   
  $m_V\sim 1\,$GeV. In this case  NP effects 
  on the luminosity measurement  will translate into  
    a large contribution to the hadronic cross section  as derived from the KLOE data.
On the other hand, for the other experiments the corresponding effects are off-resonance, and we have explicitly checked that they are negligible for the relevant parameter space of our model.
\begin{figure}[tb]
    \centering
    \includegraphics[width=0.8\textwidth]{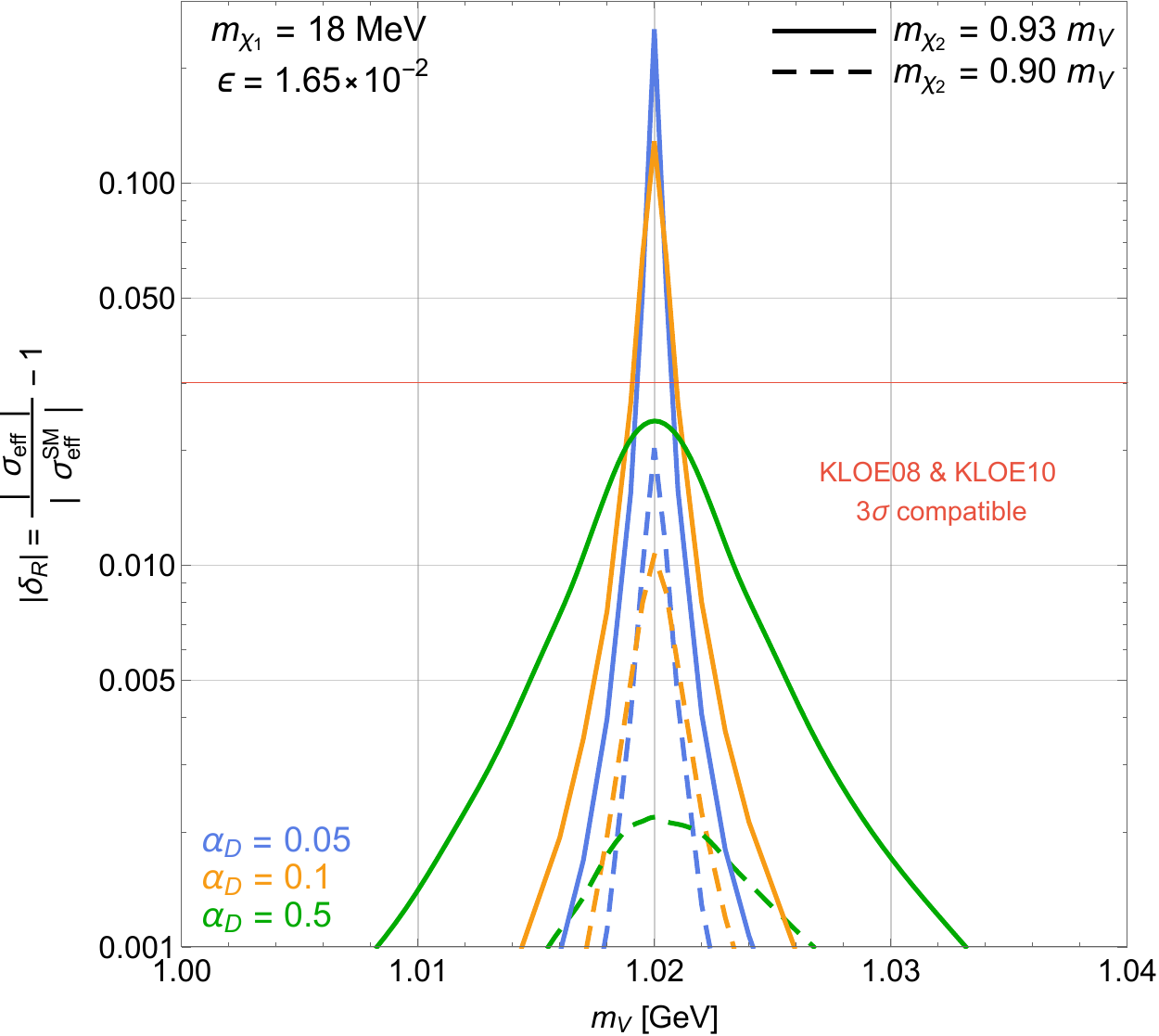}
    \caption{Relative enhancement $\delta_R$ of the inferred Bhabha cross section as a function of $m_V$ for $m_{\chi_2} = 0.93\, m_V$ (solid curves) and $m_{\chi_2} = 0.9\, m_V$ (dashed curves), and $\alpha_D = 0.05$ (blue), $0.1$ (orange) and $0.5$ (green). In this plot $m_{\chi_1} = 18$ MeV and $\varepsilon = 1.65\times 10^{-2}$. Below the red horizontal line all the KLOE measurements are consistent at a 3$\sigma$ level.}
    \label{fig:CSratioVSmV}
\end{figure}
 Figure \ref{fig:CSratioVSmV} shows the relative enhancement $\delta_R$ of \eqn{eq:deltaR}
 of the inferred Bhabha cross section as a function of the mass of the dark photon.
 This directly translates into an equal enhancement of the hadronic cross section. The dashed curves are obtained for $m_{\chi_2} = 0.90\, m_V$, while the solid curves for $m_{\chi_2} = 0.93\, m_V$. The different colors denote different values of the dark coupling $\alpha_D = 0.05$ (blue), $0.1$ (yellow) and $0.5$ (green). In addition, we fixed $m_{\chi_1} = 18 \,\rm MeV$ and the dark photon couplings to electrons and muons to $\varepsilon\, e= 5\times 10^{-3}$. As we will discuss in section~\ref{sec:constraints}, this value for the couplings is allowed by current experimental 
 results.   The red horizonatal line indicates the value of $\delta_R$ below which 
 KLOE08 and KLOE12 remain compatible  within $3\sigma$  with KLOE10.
 In this region, all other measurements also agree with  
 the combined KLOE result at better than $3\sigma$.

\subsection{Numerical analysis: the $\sigma(\mu\mu\gamma)$ method}
\label{sec:numericsmumu}

New physics strongly modifies the $\mu \mu \gamma$ cross-section without need of tuning  $m_V$ around the experimental CoM energy. This is due to the fact that the $\mu \mu \gamma$ cross-section is 
smaller than the Bhabha one.
The experimental cuts are as follow:
\begin{itemize}
    \item KLOE12~\cite{KLOE:2012anl} analysis relied on a final state with a \textit{missing photon}, with the kinematic cuts on the muons $\cos \theta_\mu \in [-0.64,0.64]$, and $p_{\mu T} \geq 160 \, \rm MeV$ or $p_{z \mu} \geq 90$ MeV and a cut on the (assumed) polar angle of the missing photon (reconstructed from the muons momenta) $\cos \theta_\gamma >15^o$. Finally, the reconstructed track mass $m_{tr}$ of each $\mu$, as defined in~\cite{KLOE:2012anl} must satisfy $m_{tr} \subset [0.08,0.115]$ to distinguish them from pions. 
    \item BESIII~\cite{BESIII:2015equ} analysis uses a final with a \textit{visible} photon, with  $|\cos \theta_\gamma | < 0.93 $ and $E_\gamma = 0.4$ GeV. Both muons are required to have $\cos \theta_\mu \in [-0.92,0.92]$, and $p_{\mu T} \geq 300 \, \rm MeV$. An additional criterium of consistency of the final states with a $\mu \mu \gamma$ is implemented via the so-called $4C$ kinematical fit. The latter cannot be straightforwardly reproduced, we note however that the cut on $\chi^2$ used in~\cite{KLOE:2012anl} $\chi^2 < 60$ is relatively loose. Accounting for the energy resolution of $2.5 \%$, we required the missing energy in the system to be smaller than $0.3$ GeV. Note that tightening this requirement to $0.2$ GeV reduces the NP contribution by three.
    \item At BaBar~\cite{BaBar:2012bdw}, both visible and invisible channels where used and the analysis is significantly more involved. In view of the importance of properly implementing the experimental cuts in the two previous experiments, we leave for future work a complete simulation of the BaBar measurement. 
\end{itemize}

We have used the same approach than for the Bhabha cross-section, relying on the \amc\ platform~\cite{Alwall:2014hca} to simulate the NP contributions. In the KLOE12~\cite{KLOE:2012anl} case, the relevant process is $e^+ e^- \to \chi_1 \chi_2$ via the resonant $V$ production. In BESIII, we instead produced directly the $V$ in addition to the hard photon $e^+ e^- \to V \gamma$, closely mimicking the SM ISR process. In both cases, the NP does not include soft or collinear photons (since a hard photon is required for BESIII) and is performed at leading order in QED.
In both cases, the amount of missing energy in our final states  determines the efficiency of the experimental cuts. In term of our iDM model, the mass $m_{\chi_1}$ of the dark matter candidate has  a strong influence on the final result, with smaller values leading to stronger effects. 

\begin{figure}[t!]
    \centering
        \includegraphics[width=0.8\textwidth]{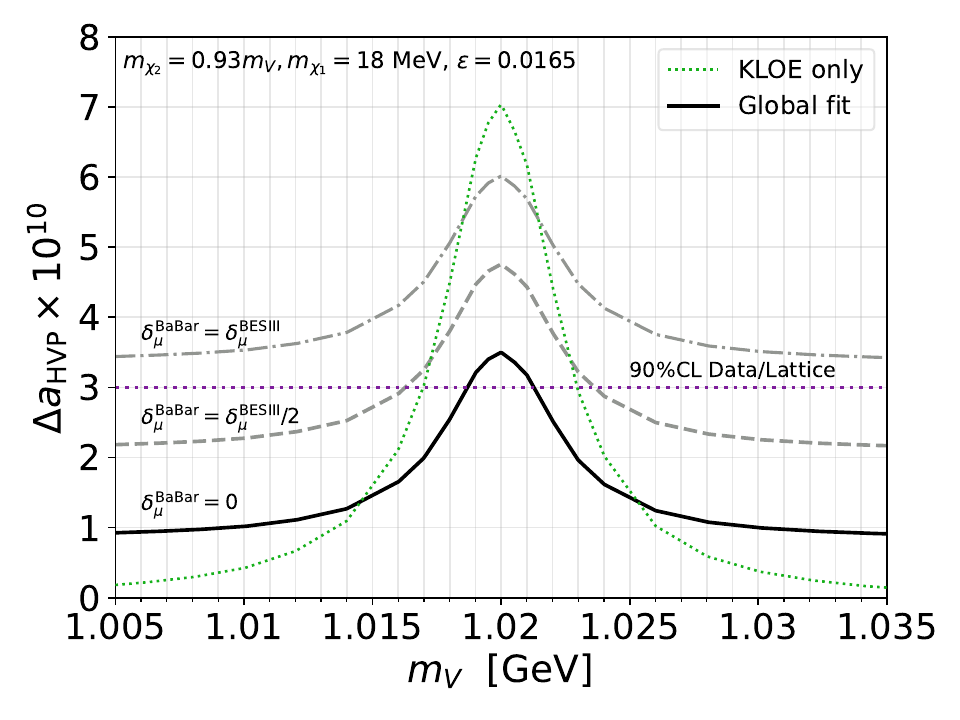}
    \caption{Theoretical prediction for the shift in $\amhvplo$ as a function of $m_V$ for a dark photon model with $m_{\chi_1} = 18$ MeV, $m_{\chi_2} = 0.93\,m_V$, $\alpha_D=0.5$ and $\epsilon=0.0165$, which leads to $\delta_\mu^{\rm KLOE12} \sim \delta_\mu^{\rm BESIII}$. 
    The dotted green line shows the combined KLOE result, while the thick black curve
    includes all the measurements, assuming  negligible shifts on BaBar results. 
    The dashed grey lines indicate the effect of an hypothetical shift in BaBar $\delta_\mu^{\rm BaBar} = \delta_\mu^{\rm BESIII}$ (top) and  $\delta_\mu^{\rm BaBar} = \delta_\mu^{\rm BESIII}/2$ (bottom). Above the dashed purple line  
    the BMW lattice result and the data-driven analysis agree at better than the $90\%$ C.L.} 
    \label{fig:deltaHVPshift}
\end{figure}

We show in Fig.~\ref{fig:deltaHVPshift} the resulting shift for a parameter point chosen for simplicity such that the relative shift in KLOE12 (denoted by $\delta_\mu^{\rm KLOE12}$), BESIII ($\delta_\mu^{\rm BESIII}$) and in KLOE08 ($\delta_R$) are of the  same order. We also include three curves depicting different guesses for the effect in BaBar 
(solid black:  $\delta_\mu^{\rm BaBar}= 0$, dashed gray:  $\delta_\mu^{\rm BaBar}= \delta_\mu^{\rm BESIII}/2$, dot-dashed gray: $\delta_\mu^{\rm BaBar}=\delta_\mu^{\rm BESIII}$). The dotted green curve shows the combined KLOE result. 
In all cases, above the dotted purple horizontal line, 
agreement between the data-driven approach and the BMW lattice result is achieved with 90$\%$ C.L.. 
Yet, the overall effect of the indirect corrections does not suffice to
provide a complete solution  to  the $\Delta a_\mu$ anomaly \eqn{eq:Deltamu}.

Let us close this section by pointing that each of the effects described above requires a different degrees of tuning on the parameters of our model. The shift  from $\sigma(\mu\mu\gamma)$ method with a visible $\gamma$ like in the analysis of ref.~\cite{BESIII:2015equ} does not require any particular tuning, and will thus be affected in any iDM model with sufficiently large kinetic coupling $\varepsilon$. (This can be seen from the constant value reached 
by  $\amhvplo$ in Fig.~\ref{fig:deltaHVPshift} when 
the KLOE CoM does not correspond precisely to $m_V$.)
The effect on  KLOE12  that exploits  the $\sigma(\mu\mu\gamma)$ method where  the photon is not detected requires $m_V \sim m_\phi$, but does not depends critically on the value of $m_{\chi_2}$. Finally, the luminosity method used by KLOE08 requires both  $m_V \sim m_\phi$ and $m_{\chi_2}\lesssim m_V$.

\section{Relevant Constraints }
\label{sec:constraints} 

In this section we study the most relevant constraints on the iDM scenario 
outlined in section~\ref{sec:model}.

\subsection{BaBar dark photon searches.} 
\noindent
The BaBar experiment made two searches relevant to our case. The first one focuses on resonances in the $e^+e^-$ spectrum with an energetic initial state radiation (ISR) photon~\cite{Lees:2014xha}. They applied the following selection cuts (with all variables in the CoM frame): $\cos\,\theta_{e^+} >-0.5$, $\cos\,\theta_{e^-}  < 0.5$, $E_{\gamma} > 0.2$ GeV and $E_{\gamma} + E_{e^+} +E_{e^-} \simeq 10.58$ GeV.

The polar angles are defined with respect to the electron beam direction, and the last selection cut requires the centre-of-mass energy of the candidate event to be within the beam energy spread, that  is around $5-10$ MeV~\cite{TheBABAR:2013jta} (see also~\cite{Seeman:2002xa}, which quotes the spread $5.5$ ($2.7$) MeV for the $e^-$ ($e^+$) beam). This search is ineffective in constraining the iDM 
model because the emission of two $\chi_1$ states always results in missing energy larger than $20$ MeV, that is larger than the beam energy spread. 

In the second search, BaBar analysed single photon events with large missing energy searching for  bumps in the photon spectrum~\cite{BaBar:2017tiz} (which superseded the older search for invisible final states in single-photon decays employed in  $\Upsilon(1S)$~\cite{delAmoSanchez:2010ac}). We reinterpret the mono-photon sensitivity 
exploiting the same strategy of \cite{Mohlabeng:2019vrz,Duerr:2019dmv,Duerr:2020muu}.  Since this search does not use cuts on simple quantities but a multivariate analysis, we use the following cuts: the events satisfy the mono-photon selection criteria if $E^\gamma_{CMS} > 2$ GeV, $-0.16 \lesssim \cos(\theta_{\gamma}) \lesssim 0.84$ and $p(e^-) < 150$ MeV or $p(e^+) < 150$ MeV. The expected sensitivity in terms of the kinetic mixing $\varepsilon = g_{eV}/ e$ for the iDM model is obtained as \cite{Duerr:2019dmv}
 \begin{equation}
 \varepsilon_{\mathrm{exp}}^{\mathrm{iDM}}
 =  \varepsilon_{\mathrm{exp}}^{\mathrm{mono}-\gamma} \sqrt{\frac{N_{\mathrm{\gamma\,\,cuts}}}{N_{\mathrm{all}}}}\,,
 \end{equation}
 where $\varepsilon_{\mathrm{exp}}^{\mathrm{mono}-\gamma}$ is the current bound  from the BaBar mono-photon analysis~\cite{BaBar:2017tiz}, $N_{\mathrm{\gamma\,\,cuts}}$ are the number of events selected using only the photon cuts, and $N_{\mathrm{all}}$ are the number of events that pass all the cuts.

Although BaBar results do not constrain significantly our scenario, the semi-visible final states could be probed in the future by Belle-II~\cite{Kou:2018nap,Duerr:2019dmv}, possibly from longer decay chain involving also the dark Higgs boson~\cite{Dreyer:2021aqd}. One key factor compared to the  analysis in Ref.~\cite{Duerr:2019dmv} is that here we are concerned with a parameter space region in which  the $V$ decay does not lead to a displaced vertex. A complete background study is thus required  to derive a proper projection of the foreseeable constraints  on our scenario.

\subsection{Effects on the $\phi$-meson  properties.}
\noindent
Since the dark photon mass must be close to the KLOE CoM energy, it will also be close to  the $\phi$-meson mass,  and thus the two widths $\Gamma_\phi$  and $\Gamma_V$ will overlap,  possibly affecting  the properties of the $\phi$. 
In particular, if the dark photon couples 
universally to all SM fermions  proportionally 
to their electric charge, then its coupling to the
strange quark will unavoidably induce $\phi-V$ mixing
(instead, in the case of a more specific lepto-philic vector boson, this would not occur, and the analysis below would not apply).

The mixed mass term  $\Mfv$ can be derived from the real part of the off-diagonal self-energy of the $(V,\phi)$ system. Assuming  $m_V \sim m_\phi$ we have
\begin{align}
    \Mfv&= \frac{1}{2m_\phi} \langle V | \frac{e \varepsilon}{3}  V_{\rho}  \bar s \gamma^\rho s  | \phi \rangle \\
        &= f_\phi \frac{e \varepsilon}{6} \sim 0.25 \, \textrm{MeV} \times \left(\frac{\varepsilon}{0.01} \right) \,,
\end{align}
where $f_\phi$ is the $\phi$ meson decay constant
defined as $$ \langle 0 | \bar s \gamma^\rho s  | \phi \rangle =   f_{\phi} m_\phi \epsilon^{* \,\rho}\ , $$ 
and we have factorised the amplitude and used
the fact that mixing can only occur between states with the same polarisation $\epsilon^*$. The evolution in time of the mixed states is governed  by the time-dependent Heisenberg equation:
\begin{align}
\label{eq:Heisen}
   i \frac{\partial }{\partial t} \begin{pmatrix}
     | \phi  \rangle \\
     | V  \rangle
    \end{pmatrix} = 
        \begin{pmatrix}
     m_\phi - i \frac{\Gamma_\phi}{2} & \Mfv \\
     \Mfv & m_V - i \frac{ \Gamma_V }{2}\\
    \end{pmatrix}
    \begin{pmatrix}
     | \phi  \rangle \\
     | V  \rangle
    \end{pmatrix}\ .
\end{align}
Mixing effects are strongly suppressed when the diagonal entries 
in the effective Hamiltonian differ sizeably, that is 
when  $|m_\phi - m_V| \gg \Mfv$ or 
$|\Gamma_\phi- \Gamma_V | \gg \Mfv$. 
For instance, in the case $m_V \simeq m_\phi$ but  $\Gamma_V \neq \Gamma_\phi$,
the mass difference $m_{\chi_2} - m_{\chi_1}$ between the two mass eigenstates is given by
\begin{align}
    m_{\chi_2} - m_{\chi_1} = \frac{2 \pi  \alpha_{\rm em} (m_\phi - m_V) \varepsilon^2 f_\phi^2}{9 (\Gamma_V-\Gamma_\phi )^2} \ .
\end{align}

For the range of parameters relevant for this work, the mass shift is both smaller than the experimental uncertainty on the $\phi$ mass ($m_\phi^{\rm exp} = 1019.461 \, \pm \, 0.016  \, \rm MeV$) and orders of magnitude lower than the uncertainty on the lattice prediction  ($m_\phi^{\rm th} = 1018 \, \pm \, 17  \, \rm MeV$~\cite{Chen:2020qma}).\footnote{The mixing is maximal when $m_V=m_\phi$ and $\Gamma_V = \Gamma_\phi$, in which case  the mass splitting is simply  $\delta m  = 2 \Mfv$. 
This shift is still an order of magnitude lower 
than the theoretical estimate for  $m_\phi$~\cite{Chen:2020qma} so that we cannot 
derive useful constraints from the mass shift. 
Note that the splitting is also smaller than  $\Gamma_\phi$, so that at  KLOE  both $\phi$ and $V$ can be  simultaneously produced on resonance.} Similarly, the shifts in the width 
of the mass eigenstates  are  safely  suppressed by $\varepsilon^2$.

Another point that we need to check concerns the 
partial width for $\phi$ decays into leptons, 
that is strongly suppressed with respect to 
hadronic decay channels, and which could be affected 
by mixing with the $V$.
The experimental value is 
$ \Gamma^{\rm exp}_{\phi ee} = 1.27 \pm 0.04 \, \rm keV$ ~\cite{ParticleDataGroup:2020ssz}
while  recent QCD lattice estimate have a theoretical errors of the same order,
dominated by the theoretical error on the $\phi$ meson decay constant $f_\phi = 241 \pm 9 \pm 2 \, \rm MeV$~\cite{Chen:2020qma}.
It is straightforward to  solve
numerically Eq.~\eqref{eq:Heisen} and derive the time-dependent evolution of the mixed states: $ | \phi (t)   \rangle= a(t)  | \phi  \rangle  + b(t)  | V  \rangle $, where the kets in the right-hand side  corresponds to the states at $t=0$. 
Since both the mixing term $b$ and  the $V$ decay width in $e^+ e^-$ are $\varepsilon$-suppressed,  the dominant modification to the $\phi $ leptonic width will arise from the time-integration of the $|a(t)|$ factor. By taking $m_V \simeq m_\phi$ and $\Gamma_{V}\neq \Gamma_{\phi}$
the change in the leptonic  width is well reproduced  by the following scaling
\begin{align}
\frac{|\delta \Gamma_{\phi e e}|}{\Gamma_{\phi e e}} \sim  \, \varepsilon^2  \frac{f_\phi}{|\Gamma_{\phi} - \Gamma_{V}|} 
\,. 
\end{align}
For all relevant points in our parameter space, this correction  remains safely below the experimental and theoretical uncertainties which are 
both of a few percent.\footnote{We have cross-checked numerically this result by adding also the
direct $V \to e^+ e^-$ contribution. In this case both the $\phi \to e^+ e^-$ and $V \to e^+ e^-$  amplitudes must be estimated and their interference must be included in the time-integrated decay rate. We have found that the 
correction saturates the 
experimental uncertainty  only for rather large 
values of  kinetic mixing, exceeding  $\sim 0.05$.}

Altogether, we can conclude that  both 
 the measurements and  the theoretical predictions for  $\phi$-meson related observables 
do not have a sufficient level of precision to
constrain effectively our NP
scenario.  
Eventually, the main  reason underlying   this conclusion  is that both the $\phi$ and the $V$ boson have very suppressed leptonic branching ratios,
while their main decay channels $\phi \to \mathrm{hadrons}$
and $V\to \chi_1\chi_2$ are  completely different final states. 

\bigskip
\subsection{KLOE10 off-resonance measurement}
\label{sec:KLOE off-res}

 The second analysis from the KLOE collaboration,  KLOE10~\cite{KLOE:2010qei}, was performed $\sim 20\,$MeV 
 below the $\phi$ resonance, at $\sqrt{s} = 1.00$ GeV. This implies that if the $V$ width is significantly smaller than $20$ MeV, 
 KLOE08~\cite{KLOE:2008fmq} and KLOE10~\cite{KLOE:2010qei} measurements 
 cannot be simultaneously shifted by similar amounts.
A naive comparison indicates that the two measurements would remain in agreement at the $2\sigma$ level  
if the shift does not exceed  $\delta_R \simeq 2 \%$. However, the final KLOE result is in fact a simultaneous fit to  
three analysis (including the last KLOE12~\cite{KLOE:2012anl}) which was carried out by the KLOE collaboration 
in Ref.~\cite{Anastasi:2017eio}. The combined KLOE result takes into account sizeable correlations
between various systematic uncertainties  which dominate the overall error, and which increase it by about $60\%$
with respect to a naive estimate assuming uncorrelated measurements.\footnote{KLOE08 and KLOE12 have also a fully correlated statistical error, since are based on the same dataset.  
Statistical errors are, however, a subdominant source of uncertainty.} 
Systematic correlations might thus allow for somewhat larger 
values of $\delta_R$ while maintaining compatibility between the different results, and  
for this reason we relax the compatibility requirement to 
 \begin{align}
     \delta_R \lesssim 2.5 \% \ \ \textrm{ when } \Gamma_{V} \ll 20 \ \rm MeV\,.
 \end{align}
 In our benchmark iDM scenario, the $V$ width is typically of the order of few MeVs, making this constraint relevant. 
 However, this is still a large enough shift to solve the tension with the BaBar result.

\bigskip
\subsection{Muon cross-section measurements}
\label{sec:KLOEmuon}

The KLOE~\cite{KLOE:2012anl}, BaBar~\cite{BaBar:2012bdw} and BESIII~\cite{BESIII:2015equ} collaborations 
have also measured  $\mu \mu \gamma$ final states to determine the differential luminosity 
by comparing with the theoretical $\sigma_{\mu\mu}\equiv \sigma(e^+e^- \to \mu^+ \mu^- \gamma)$ cross-section,  
and cross-checked the results with the total luminosity as determined from Bhabha scattering. 
Agreement is usually obtained at or below the percent level for the overall integrated cross-sections. However, 
for specific subsets of  di-muon invariant mass bins deviations can be significant 
(see e.g. the  data in the $m_{\mu\mu}^2 \sim (0.6-0.9)\,$GeV$^2$  windows reported in Fig.\,5 in Ref.~\cite{KLOE:2012anl}, Fig.\,32 in Ref.~\cite{BaBar:2012bdw} and Fig.\,1 in Ref.~\cite{BESIII:2015equ}). 
We will focus here on the  analysis of the KLOE collaboration~\cite{KLOE:2012anl} (KLOE12) which exploits $\mu \mu \gamma$ final states in which the $\gamma$ is not reconstructed. Assuming that the branching fractions for $V$ boson decays into muons and electrons
are similar, as would be implied by universality of the $V$ couplings to leptons, we 
would expect a  relative excess of di-muon events similar to the excess of $e^+e^-$  Bhabha events in the KLOE08  measurement~\cite{KLOE:2008fmq}. KLOE12 further observed that the integrated $\mu \mu \gamma$ cross-section was in accordance with the theoretical prediction up to a $\sim 1\%$ systematic uncertainty. Since this 
measurement relied on the same method to estimate the luminosity~\cite{KLOE:2006itf} as the KLOE08 analysis, 
in reconstructing the cross-section the direct contribution to $\mu\mu$ events from $\mu^+ \mu^- (\chi_1 \chi_1)$ 
can be compensated at least in part by the effect of overestimating the luminosity because  of a similar excess 
of $e^+ e^- (\chi_1 \chi_1)$ Bhabha events. 
In fact, let us consider Fig. 5 in Ref.~\cite{KLOE:2012anl} 
where the  number of $\mu\mu\gamma$ events expected in the SM  $N^{SM}_{\mu \mu}$ was estimated with the Phokhara MC,
and compared  to the experimental data $N^{exp}_{\mu\mu}$. From the QED cross-section $\sigma_{\mu\mu}$ 
one estimates 
\begin{align}
\label{eq:NSM}
 N^{SM}_{\mu \mu} = \epsilon^{SM}_\mu \sigma_{\mu \mu}  \times  \mathcal{L}^{\rm SM}_{e^+e^-}
\end{align}
where $\epsilon^{SM}_\mu$ is the experimental efficiency for the SM signal and  
$\mathcal{L}^{\rm SM}_{e^+e^-} = \mathcal{L}_{e^+e^-} (1+\delta_R)$ is the (overestimated)  luminosity inferred by assuming 
only QED Bhabha scattering, while $ \mathcal{L}_{e^+e^-}$ is the true luminosity obtained after 
accounting for the NP contribution $\delta_R$ to $e^+e^-$ events, see \eqn{eq:lum} in Sec.~\ref{sec:mod_lumi}.
On the other hand, the  experimental data include also a direct NP contribution $ \sigma_{\mu \mu X}^{NP}$ from 
$\mu^+ \mu^- (\chi_1 \chi_1)$ events mimicking  $\mu \mu \gamma$  final states with an undetected $\gamma$, and 
measured with  efficiency  $\epsilon^{NP}_\mu $, so that 
\begin{align}
\label{eq:Nexp}
 N^{exp}_{\mu \mu} = \left(\epsilon^{SM}_\mu \sigma_{\mu \mu}  + \epsilon^{NP}_\mu  \sigma_{\mu \mu X}^{NP}\right)\times \mathcal{L}_{e^+e^-} 
 = 
 \epsilon^{SM}_\mu \sigma_{\mu \mu} \left(1+ \delta_\mu\right) \times \mathcal{L}_{e^+e^-}
 \ .
\end{align}
All in all, from equations \ref{eq:NSM} and \ref{eq:Nexp}, we obtain:
\begin{align}
\label{eq:NNratio}
 \frac{N^{exp}_{\mu \mu}}{N^{SM}_{\mu \mu}} = \frac{1+\delta_\mu}{1+\delta_R } \,,
\end{align}
which shows how the direct effect $\delta_\mu$ and the indirect (luminosity-related) effect $\delta_R$ 
tend to compensate in the ratio. 
Note, however, that while  the relative luminosity shift $\delta_R$ 
defined in Sec.~\ref{sec:mod_lumi} is independent of the CoM energy of the $e^+e^-$ collision, that is,   
it remains constant with respect to the reconstructed  di-muon invariant mass squared $s'$, 
this is not true for the direct effect. This depends on the different energy behaviour of the SM and NP 
cross sections, so that we have as in Eq.~\eqref{eq:dmu}:
\begin{align}
\delta_\mu (s')~=~ \frac{\sigma^{NP}_{\mu\mu X}  (s') }{ \sigma_{\mu\mu}  (s')}  \frac{\epsilon^{NP}}{\epsilon^{SM}}\ ,
\end{align}
with the experimental efficiencies explicitly included for clarity. 
As a consequence of the $s'$ dependence in $\delta_\mu$, also the ratio in \eqn{eq:NNratio} will depend on $s'$. 

In order to numerically estimate the predicted ratio ${N^{exp}_{\mu \mu}}/{N^{SM}_{\mu \mu}}$
with the NP effects included,  we have implemented in the \amc\ platform~\cite{Alwall:2014hca} the $e^+e^- \to \mu^+ \mu^- \chi_1 \chi_1$ process along with the cuts described in Ref.~\cite{KLOE:2012anl}. In particular, at KLOE the identification of 
 $\mu^\pm$ (and $\pi^\pm$) events relies on the ``computed track mass'' $m_{tr}$, which is defined  in terms the momentum $p_+$ and $p_-$ of the reconstructed positively and negatively charged tracks, and of the missing energy of the event,  as the solution of  energy conservation.  Under the assumption that the missing energy is carried away by a photon we have $E^2_\gamma =|\vec{p}_\gamma|^2 = 
|\vec{p}_- + \vec{p}_+|^2 $, and the energy conservation condition reads
\begin{align}
\label{eq:mtr}
    (\sqrt{s}-\sqrt{|\vec{p}_+|^2+m^2_{tr}}-\sqrt{|\vec{p}_-|^2+m^2_{tr}})^2 -|\vec{p}_- + \vec{p}_+|^2 = 0 \ .
\end{align}
For muon identification it is required that the solution  satisfies $80 \leq m_{tr}/\mathrm{MeV} \leq 115\,$.  
However, for $\mu^+ \mu^- (\chi_1 \chi'_1)$ final states energy conservation would imply the replacement 
$|\vec{p}_- + \vec{p}_+|^2 \to (E_{\chi_1}+E_{\chi'_1})^2 $ so that 
if \eqn{eq:mtr} is instead still used, one tends to obtain  track mass solutions that are somewhat larger than the muon mass 
(with values that depend on $M_{\chi_1}$), so that certain $\mu^+\mu^-$ events do not pass the $m_{tr}$ cut and are rejected. 
However, the most important suppression of the number of  $\mu^+ \mu^- (\chi_1 \chi'_1)$ events that pass the cuts 
arises from the requirement that the polar angle $\theta_{\mu\mu}$ of the di-muon  
momentum $\vec{P}_{\mu\mu} = \vec{p}_+ + \vec{p}_-$ satisfies $\cos \theta_{\mu\mu} > \cos 15^\circ$.\footnote{This cut selects 
recoiling photons that are emitted at small angles with respect to the beam direction, 
which largely enhances the number of ISR with respect to FSR $\gamma$ events.}
It is thus clear that if we assume that   $ \chi_1 \chi'_1$ are emitted approximately isotropically, 
the momentum of the recoiling di-muon system will pass this cut only a fraction $(\sin15^\circ)^2/2\sim 3.5\%$ of the times.

\begin{figure}[t!]
    \centering
        \includegraphics[width=0.8\textwidth]{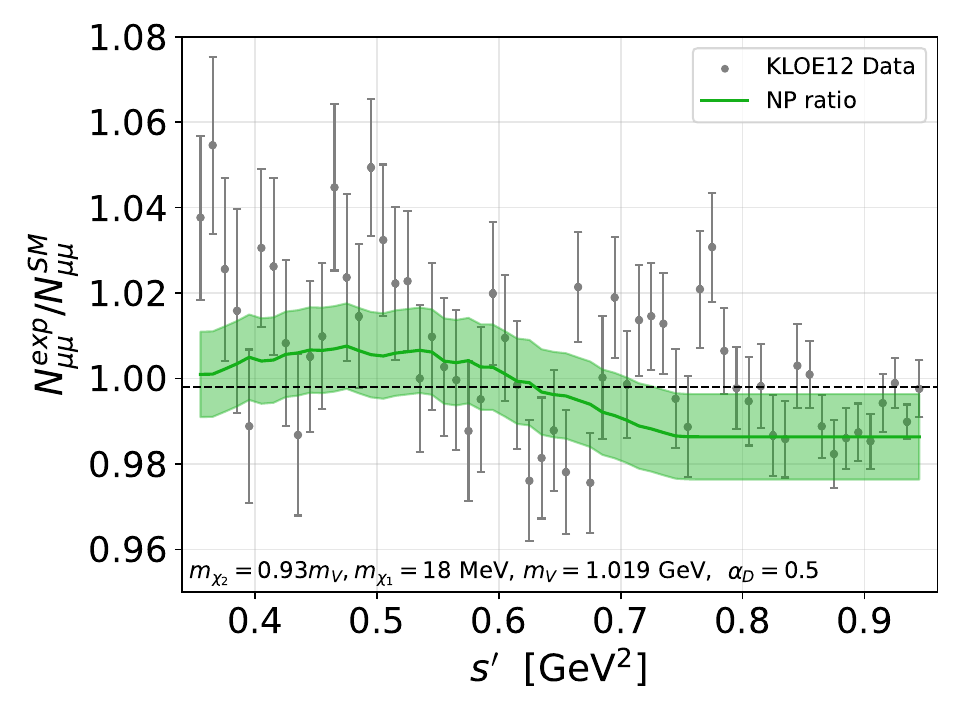}
    \caption{
    New physics contributions on the ratio $N^{exp}_{\mu \mu}/N^{SM}_{\mu \mu}$  for our iDM model with $m_{\chi_2}=0.95\,\textrm{ GeV}, m_{\chi_1}= 18\,\textrm{ MeV}$ and $\varepsilon = 0.013$ (green line). 
    The grey points are the KLOE data~\cite{KLOE:2012anl} and the black line the data fit to a constant ratio. We have added a 1\% uncertainty corresponding to the experimental systematics for visualisation~\cite{KLOE:2012anl}.}
    \label{fig:deltamu}
\end{figure}

We illustrate the results of our simulation in Fig.~\ref{fig:deltamu}. The slope of the ratio is generated by the  
$\delta_\mu(s')$ correction and is mostly active at small values of  $s'$, where the SM $\mu \mu \gamma$ signal 
is more suppressed. The luminosity effect instead shifts the entire curve downwards. 
All in all, we have estimated the $\chi^2$ value for the green curve in Fig.~\ref{fig:deltamu}, 
and we have found that it yields a value roughly similar to the $\chi^2$ of the constant fit.\footnote{A more recent dataset on muonic events with similar errors was presented in Ref.~\cite{KLOE-2:2016mgi}.}
However, the NP effect has the interesting feature of being able to better accommodate, at least qualitatively, 
the behaviour of the data, which  at small $s'$  overshoot the constant fit,  while  for large $s'$ undershoot it.

\subsection{KLOE forward-backward asymmetry.}
\label{sec:Afb}
The KLOE collaboration performed an analysis of the forward-backward asymmetry $A_{\rm FB}$ in  $e^+e^- \to e^+e^-$
around the $\phi$ resonance 
in order to extract the leptonic widths 
$\Gamma_{\phi ee}$ and $\Gamma_{\phi \mu\mu}$
and to carry out a test of lepton flavour universality~\cite{KLOE:2004uzx}. 
They measured $A_{\rm FB}$ for three different CoM energies:
\begin{align}
\label{eq:KLOEAfb}
    A_{\rm FB} (\sqrt{s}) = \begin{cases}
    0.6275 \pm 0.0003 \quad (\sqrt{s} = 1017.17 \, {\rm MeV }\simeq m_\phi - \Gamma_\phi/2  ) \\
        0.6205 \pm 0.0003 \quad  (\sqrt{s} = 1019.72 \, {\rm MeV} \simeq m_\phi ) \\
            0.6161 \pm 0.0004 \quad (\sqrt{s} = 1022.17  \, {\rm MeV} \simeq m_\phi + \Gamma_\phi/2  ) \ ,
    \end{cases}
\end{align}
where all three data-points share a common systematic uncertainty of $0.002$. 
The collaboration then fitted these three experimental values to the interference pattern expected from the $\phi$ to obtain the measurement $\Gamma_{\phi ee} = 1.32 \pm 0.05 \pm 0.03$ keV. An off-shell $V$ exchange will also induce an interference pattern in the asymmetry, we can re-interpret this measurement as being a limit on the $V$ contribution.  However, we also must include the direct contribution from the $e^+ e^- \to V \to \chi_1 \chi_1 e^+ e^- $ process. 
 Let us denote  the measured  number of forward/backward events as $N_F$ and $N_B$. The asymmetry can be decomposed as
 \begin{align}
 A_{\rm FB} (\sqrt{s}) ~&\equiv~ \frac{N_F - N_B}{N_F +N_B}   \\
 & =  A_{\rm FB}^{\rm Bhabha} \times ( 1 - \delta_R(\sqrt{s})) + A_{\rm FB}^{\phi } + A_{\rm FB}^{V } \ .
 \end{align}
The first term of the second line includes a correction 
which is due to the fact that the process $e^+ e^- \to V \to \chi_1 \chi_1 e^+ e^- $  contributes to the denominator of the Bhabha asymmetry while, since it has a  negligible asymmetry, 
it does not contribute to the numerator.  The second and third contributions correspond to  the interference between 
virtual photon exchange and respectively the $\phi$ and the $V$ vector boson. Since the $V$ is produced resonantly, the correction $\delta_R$ depends strongly on $\sqrt{s}$ and on the  value of the $V$ width. The effect of $\delta_R$ is to reduce the asymmetry, while both the interference terms contribute positively. 

In order to assess the complete $V$ contribution, we have simulated the full process $e^+ e^- \to e^+ e^-$ in  \amc, implementing the cuts given in~\cite{KLOE:2004uzx}.  It 
 is clear that the fit performed by the 
 collaboration  cannot be applied to the scenario at hand  with multiple interference terms plus the ``inverse resonance'' effect. Hence we instead adopt a more conservative proxy, the difference between the lowest and the highest measurements of $A_{\rm FB}$ in $\sqrt{s}$:
\begin{align}
\label{eq:asymdef}
    \Delta A_{\rm FB} &~\equiv~ \frac{A_{\rm FB} ( m_\phi - \Gamma_\phi/2) -     A_{\rm FB} ( m_\phi + \Gamma_\phi/2)}{A_{\rm FB} ( m_\phi - \Gamma_\phi/2) +     A_{\rm FB} ( m_\phi + \Gamma_\phi/2} \\
    & = \Delta A_{\rm FB}^\phi + \Delta A_{\rm FB}^V -\frac{ \delta_R(\sqrt{s_-}) -  \delta_R(\sqrt{s_+})}{2} \ ,
\end{align}
where  $s_{\pm} \simeq  m_\phi \pm \Gamma_\phi/2$ 
are the CoM energies of the KLOE measurements, cf. Eq.~\eqref{eq:KLOEAfb}.
KLOE measured $\Delta A^{\rm exp}_{\rm FB} = (9.17\pm 0.35) \cdot 10^{-3} $.
In either the vector meson dominance\footnote{See 
 e.g. Ref.~\cite{Fujiwara:1984mp} for more details on the Vector Meson Dominance approach (VMD) which has been widely used in recent dark sector literature~\cite{Tulin:2014tya,Ilten:2018crw,Darme:2020gyx} for the $\phi$ and dark photon-related amplitudes.} or in a simple factorisation approach, the $\phi$-mediated interference cross-section can be obtained as
 \begin{align}
   \sigma_{\rm int} =   \frac{3 \alpha_{\rm em} \Gamma_{\phi ee} } {m_\phi} \, \frac{s - m_\phi^2 }{ (s - m_\phi^2)^2 + s \Gamma_\phi^2} \, \int_{c_{\rm min}}^{c_{\rm max}} d c_\theta \left[ \pi  \left( c_\theta^2-\frac{(c_\theta+1)^2}{1-c_\theta} +1 \right) \right] \ , 
 \end{align}
 where $c_\theta$ is the outgoing electron angle and $c_{\rm min}, c_{\rm max}$ are either the acceptance or $0$ for the forward/backward case. Estimating the theoretical uncertainties on this expression is delicate, although it is clear that the theoretical prediction for $ \Gamma_{\phi ee}$  plays the leading role.
We can get an estimate 
of this effect by using  the 
lattice result  $f_\phi$.   
The most recent estimate~\cite{Chen:2020qma} gives  
 $f_\phi= 241 \pm 9 \pm 2 \, \rm MeV$.
However, an earlier lattice estimate~\cite{Jansen:2009hr} found $f_\phi = 308 \pm 29 \, \rm MeV$ which, in spite of the
much larger error, differs from the result of Ref.~\cite{Chen:2020qma} by more than $2\sigma$. Another 
earlier estimate~\cite{Donald:2013pea} found a central 
value  in agreement with Ref.~\cite{Chen:2020qma}
but with twice the error $f_\phi = 241 \pm 18 \, \rm MeV$. This latter uncertainty would correspond to a $ 1.8 \cdot 10^{-3}$ uncertainty on $  \Delta A_{\rm FB} $, while  
the difference between the central values of Refs.~\cite{Chen:2020qma} and \cite{Jansen:2009hr} 
would translate into an uncertainty of $\approx 6.7 \cdot 10^{-3}$.

From our simulations, we obtain a theoretical prediction for the contribution from Bhabha and the $\phi$ resonance of $  \Delta A^{\rm th}_{\rm FB} = 12.0 \cdot 10^{-3}$ which also differs from the experimental measurement by $2.8 \cdot 10^{-3}$. Altogether, it is clear that additional theoretical input on the $\phi$ interference contribution would be required to match the experimental precision. We will therefore present in the rest of this work both a conservative estimate based on the discrepancy between the lattice estimate of Refs.~\cite{Chen:2020qma} and \cite{Jansen:2009hr} (corresponding to a $\sim {}^{ +7.0 }_{- 5.5} \cdot 10^{-3}$ error), and a more aggressive limit based only on the difference between our estimate of the SM $\phi$ contribution and the measurement ($\sim \pm 3 \cdot 10^{-3}$).
If we neglect the interference term and consider the small width regime $\Gamma_V < \Gamma_\phi$, then for the case of maximum negative shift, $m_V \simeq \sqrt{s_-}$, we have $\Delta A^{\rm NP}_{\rm FB} \simeq  - \delta_R(\sqrt{s_-})/2$. 
For the ``aggressive'' limit this translates into a maximum shift on the luminosity of $\delta_R \simeq 1.2 \%$. As shown in Fig.~\ref{fig:CSratioVSmV}, this still allows to bring KLOE in agreement within $2\sigma$ with BaBar.

In practice, the interference contribution is also important and allows for somewhat larger shifts due to  cancellations between both terms, as is confirmed by the full numerical results. Furthermore, the above limit is strongly reduced for larger width with a $\Gamma_\phi^2/\Gamma_V^2$ suppression of $\Delta A^{\rm NP}_{\rm FB} $ when $\Gamma_\phi \ll \Gamma_V$.  Finally, while the analysis in Ref.~\cite{KLOE:2004uzx}  relied on a very precise calibration of the CoM energy, this was not the case for the study used to derive the $\amhvp$ KLOE contribution. As can be seen e.g. from Fig.~7 of Ref.~\cite{KLOE:2006itf}, the spread in the CoM energy is larger, of the order of MeV, and this is related  to the method  used to 
measure the hadronic cross-section via ISR. Since we are primarily interested in this measurement,  then we  need to consider an uncertainty on  $m_V$ of the same order.\footnote{This prevents us to use  the asymmetries corresponding to the $\sqrt{s_0} \simeq m_\phi$ bin,  since the two points are separated by  an energy interval of the order of the uncertainty on  $m_V$.}

Although limited by theoretical uncertainty, the forward-backward asymmetry measurement can provide significant constraints on our scenario. This mostly occurs because in the particular iDM model we have considered, the $V$ width (see  Eq.~\eqref{eq:widthV}) is always in the MeV range. It would 
certainly be interesting to have a more complete experimental dataset,  leveraging this observable to constrain this type of physics, for example from the CMD-3 experiment~\cite{Ryzhenenkov:2020vrk}.

Finally, we did not include possible limits from the value of the $\phi$ partial width into muons 
reported in the same reference~\cite{KLOE:2004uzx}. 
This is because firstly $\Gamma_{\phi\mu\mu}$ was inferred from cross-sections measurements only and thus has significantly larger experimental uncertainties, and moreover it is also  sensitive to the Bhabha luminosity shift that we have described above. Secondly, it depends on the $V$ coupling to muons, which do not directly enter 
our mechanism to shift $\amhvp$, and also on the 
mass $m_{\chi_1}$ of the lightest dark matter particle, since 
for the $\mu^+\mu^-$ channel the experimental cuts on the missing energy are significantly  tighter. Similar considerations can be applied to the estimates of $\Gamma_{\phi ee}$ derived from fitting  measurements of cross-sections into hadronic final states.

\bigskip
\subsection{Indirect effects on LEP precision measurements.} 

\noindent
A new vector boson of \mbox{$\sim 1\,$GeV mass} 
 can give rise to different types of  
indirect effects on LEP electroweak precision measurements. 
A first effect is a modification of the value of the
electromagnetic coupling constant extrapolated at the 
large CoM energy scale of the LEP $e^+e^-$ collisions, that can be written 
as 
\begin{equation}
    \label{eq:alphaMZ}
     \alpha(s) = \frac{\alpha}{1-\Delta\alpha_\ell(s)-\Delta\alpha_{\rm top}(s) - \Delta\alpha_{\rm had}^{(5)}(s)}\,,
    \end{equation}
where $\Delta\alpha_\ell$ and $\Delta\alpha_{\rm top}$ 
are the contribution to the photon vacuum polarization 
from the leptons and the top quark, which can be computed 
perturbatively with good accuracy, while the five-flavor hadronic contribution 
$\Delta\alpha_{\rm had}^{(5)}$ has to be extracted from data, and is determined by 
the dispersion relation
\begin{equation}
 \label{eq:dispersionMZ}   
\Delta\alpha_{\rm had}^{(5)}(s) = \frac{s}{4\pi^2\alpha} P\!\!\int ds' 
\frac{\sigma_{\rm had}(s')}{s-s'}\,,
\end{equation}
where $P$ denotes the integral principal value.
The important difference  with the otherwise similar 
expression in \eqn{eq:amhvp} is the $1/s$ factor in that equation 
which implies that low energy data play a dominant role for $a_\mu$,   
while e.g. for  $s\simeq M^2_Z$  the low energy contribution to the  
integral in  \eqn{eq:dispersionMZ} is much less relevant. 
Hence, an increased value of the KLOE result for $\sigma_{\rm had}$ 
in the [0.6,0.9]\,GeV energy range is unlikely to affect the 
overall agreement with the LEP electroweak precision data
(see also the dedicated analyses in Refs.~\cite{Keshavarzi:2020bfy,Crivellin:2020zul}).

Additional subtle effects are related to corrections 
to luminosity measurements. 
The importance of a reliable determination of the LEP luminosity 
is well exemplified by the recent reassessment of the LEP measurement of the 
number of light active  neutrino species $N_\nu$~\cite{Voutsinas:2019hwu}.    
During the first phase (LEP-1)  high statistics 
data were collected at and around the $Z$ pole, providing 
a wealth of measurements with sub-percent precision~\cite{ALEPH:2005ab}.
In particular, the value of  $N_\nu$ 
was estimated from the measurement of the hadronic  peak cross section 
$\sigma^{\rm peak}_{\rm had}$ by means of the relation 
\begin{equation}
    \label{eq:Nnu}
N_\nu\left(\frac{\Gamma^{\rm SM}_{\nu}}{\Gamma^{\rm SM}_\ell}\right)
= \left(\frac{12 \pi}{m^2_Z} \frac{R^0_\ell}{\sigma^{\rm peak}_{\rm had}}\right) - R^0_\ell -3 -\delta_\tau
    \end{equation}
    where $\Gamma^{\rm SM}_{\nu,\ell}$ are the SM predictions for the 
   partial $Z$ decay width into 
    neutrinos and   (massless) charged leptons, 
$R^0_\ell$ is the ratio between the $Z$ branching fractions into hadrons and leptons
and $\delta_\tau\sim 2.26 \times 10^{-3}$ accounts for a small correction 
from finite $m_\tau$ effects on the $\Gamma_\tau^{\rm SM}$ partial width. 
The combination of the measurements made by the four LEP experiments~\cite{ALEPH:2005ab}
led to: $N_\nu = 2.9840 \pm 0.0082$ which had a $2\sigma$ tension with the canonical SM
value  $N^{\rm SM}_\nu = 3$.
The measurement is directly affected by biased errors in estimating 
the integrated luminosity only through $\sigma^{\rm peak}_{\rm had}$, 
all other quantities in \eqn{eq:Nnu} 
instead do not depend on the absolute luminosity. Indeed, the uncertainty 
on the  integrated luminosity represents the largest contribution to the uncertainty on $N_\nu$. 
The LEP luminosity was determined by comparing the measured  rate of 
Bhabha-scattering process at small angles with the SM prediction.
Recently this determination has been reassessed 
first by correcting for a bias in the luminosity measurement due to 
the large charge density of the particle bunches which modify (decrease) the effective 
acceptance of the luminometer~\cite{Voutsinas:2019hwu}, and next by using an updated and 
more accurate prediction  of the Bhabha cross section  which is found to reduce its value 
by about 0.048\%~\cite{Janot:2019oyi}.
Both effects go in the direction of decreasing the Bhabha cross section 
with respect to that used in the LEP fit~\cite{ALEPH:2005ab}, 
thus increasing the effective luminosity, decreasing 
$\sigma^0_{\rm had}$, and eventually raising the inferred value of active neutrino species to  $N_\nu = 2.9963 \pm 0.0074$, in perfect  agreement with the SM. 

The direct effect of a  (constructive) NP contribution to the Bhabha cross 
section would instead go in the opposite direction, decreasing the estimated 
luminosity, and hence  increasing $\sigma^{\rm peak}_{\rm had}$ and reducing $N_\nu$. 
However,  at LEP energies the $m_V\sim1\,$GeV vector boson 
grossly behaves like a massive photon with $\varepsilon$-suppressed couplings
to electrons/positrons. The most relevant effect will then come from  $t$-channel $\gamma$-$V$ interference,  which in our case is suppressed by a relative $\varepsilon^2\sim 10^{-4}$, and thus negligible. 
An additional indirect effect is again related to 
adjustments  in   
the value of the hadronic vacuum polarization that enters  
the $t$-channel photon propagator, and that could  modify the value 
of the $\alpha(t)$ input to the LEP  Bhabha  event generators~\cite{Jadach:1996gu}. 
Yet, given that at the small Bhabha scattering 
 angles/small momentum transfer  ($\theta \lesssim 60\,$mrad, $t\lesssim (2.8\,\mathrm{GeV})^2$) specific of the LEP luminosity measurements   
  the hadronic contribution remains small 
  ($\Delta\alpha_{\rm had}^{(5)}\lesssim 0.008$ and at most $\sim 30\%$ of the total vacuum polarization, see e.g. Ref.~\cite{Jadach:1996gu})  
  the corresponding correction remains 
  below the systematic  uncertainties.
  It is interesting to note that, 
although  both the direct  contribution of $V$ to  Bhabha scattering 
and the indirect effect  of increasing the value of  
$\Delta\alpha_{\rm had}^{(5)}$ affect negligibly the LEP luminosity 
measurement, they  go in the same  direction and, for example, both 
would tend to decrease   the central value of $N_\nu$.

\subsection{LEP limit on $V-Z$ mixing} In the inelastic dark matter model, the interaction between the dark photon and the SM particles proceed via kinetic mixing between the $U(1)_V$ and the $U(1)_Y$ field strength. This implies that after electroweak symmetry breaking, the $Z$ boson mixes with the dark photon, leading to a small modification of the SM electroweak couplings. A complete fit for this effect in the electroweak precision observables was performed in~\cite{Hook:2010tw,Curtin:2014cca} leading to the relatively model independent bound:
\begin{align}
   \varepsilon < 0.027 \ \  \textrm{(LEP - EW fit)}\,.
   \label{eq:lep-ew-fit}
\end{align}

\section{Joint solution to the  $a_\mu$-related anomalies}
\label{sec:allinall}

In this section we study in which range of parameters for the model 
outlined in section~\ref{sec:model} the indirect NP-induced effects on $a_\mu$, discussed in the Sections \ref{sec:mod_lumi} and \ref{sec:mod_mumugamma},
together with the direct loop contributions of the new vector boson $V$ 
can optimally resolve the various $a_\mu$ related discrepancies.

\begin{figure}[t!]
    \centering
        \includegraphics[width=0.8\textwidth]{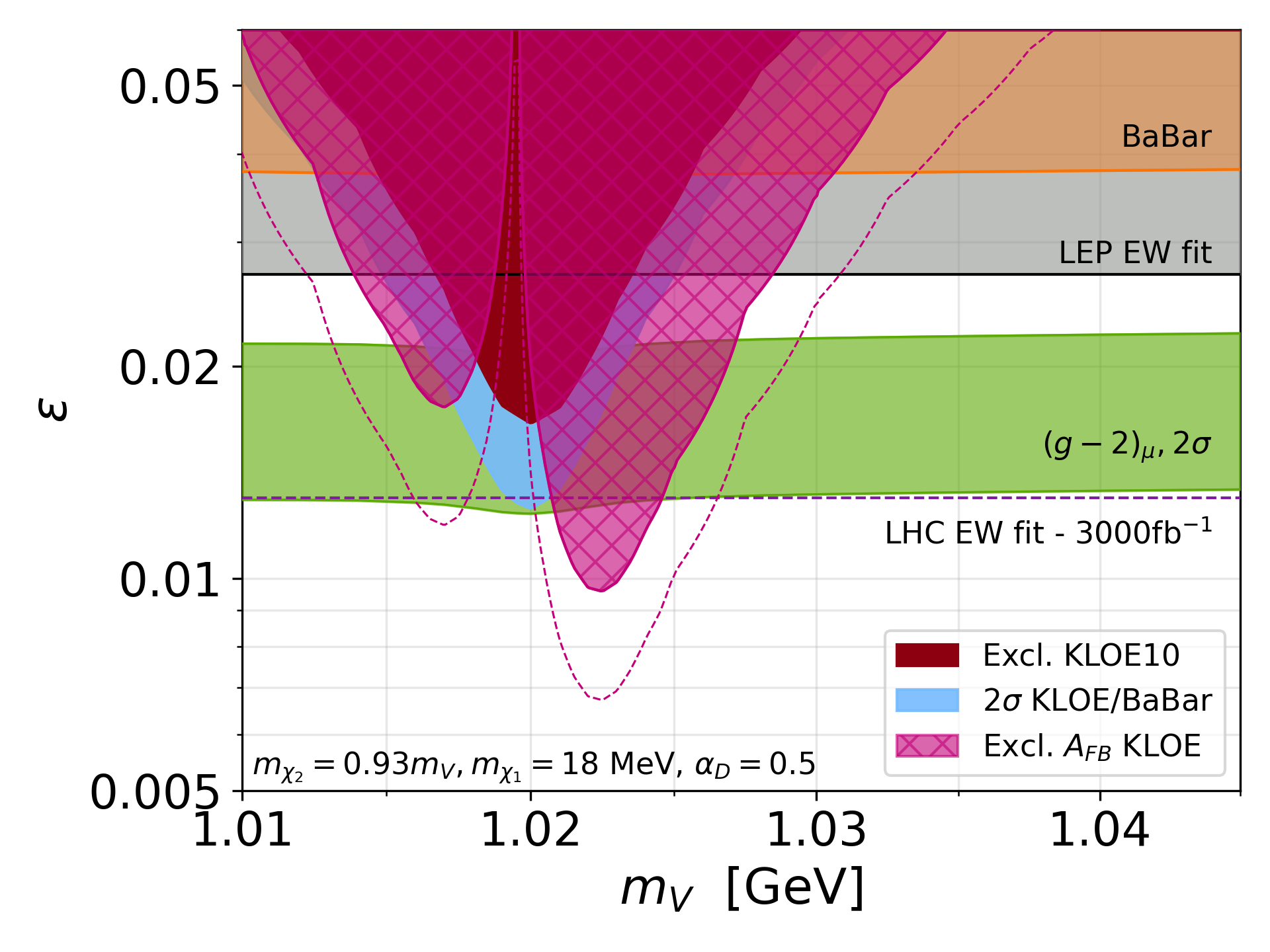}
    \caption{Parameter range compatible at $2 \sigma$ with the experimental measurement of $\damtwo$ (green region)   resulting from a redetermination of the KLOE luminosity, for  $\alpha_D=~0.5,  m_{\chi_2}=0.93~m_V $ and $m_{\chi_1}= 18 $ MeV. 
    In the blue region the KLOE and BaBar results for $\shad$  
    are brought into agreement at $2\sigma$. The red  region corresponds to a shift of the KLOE measurement in tension with BaBar (and  with the other experiments) at more than $2\sigma$. 
        The limit from the electroweak fit at LEP (gray band), the projection for  LHC  run-3~\cite{Curtin:2014cca} (violet dashed line), and the recasting of the BaBar limit~\cite{Lees:2014xha} (orange band) are also shown (see text). The hatched magenta region corresponds to the   conservative $2\sigma$ exclusion from $\Delta A_{FB}$, while the  magenta dashed line corresponds to the more  aggressive exclusion limit (see text).}
    \label{fig:MVplot}
\end{figure}

First we present in Fig.~\ref{fig:MVplot} the parameter space in which one can obtain a simultaneous fit to both the BaBar/KLOE discrepancy and the $a_\mu$ at the $2\sigma$-level. The blue region denotes the values of $\varepsilon$ where the KLOE result is within $2\sigma$ of the BaBar measurement assuming that the BaBar result is not modified, while the green region is the area where $a_\mu$ fits the experimental result at the $2\sigma$-level (with both the shift in the data-driven estimate and the new physics contribution included). As expected from the resonant nature of the production mechanism, the shift in KLOE data brings it in agreement with BaBar only in a narrow region around the $\phi$ mass. Depending on the strength of the shift, it can  account for up to one-quarter of the full anomaly before leading to new tensions with the rest of the data-driven experimental results. Conversely, our results can be interpreted as a new exclusion limit for this type of dark photons centred around the GeV (represented as the red area in the figure) for which the shift lead to $3\sigma$ internal tension between the different KLOE measurements. We note that similar exclusions likely exist around the CoM energy of the different experiments using Bhabha scattering to calibrate their luminosity. The magenta area represents the $2\sigma$ ``conservative'' exclusion using the $\Delta A_{FB}$ measurement as presented in section~\ref{sec:Afb}, while the dashed magenta line is the ``aggressive'' exclusion.

When the dark photon mass becomes close to $1\, \rm GeV$, the KLOE10~\cite{KLOE:2010qei} analysis result will received a correction similar to the one for KLOE08 at $m_V \sim 1.02$ GeV. While this circumvents efficiently all the $\phi$-meson related constraints, it also provides a significantly weaker overall effect since both the KLOE08 and KLOE12 result are not modified. It can nonetheless provide a definite improvement in the overall agreement between the data-driven method and the lattice BMW result, although with a NP contribution almost overshooting the $\amtwo$ anomaly. Finally, the shift in BESIII does not rely on a tuned $m_V$ value and will be active for all vector masses in the GeV region. If a similar effect could be quantitatively shown to exist in BaBar, it would represent a generic mechanism to reduce the discrepancies with the BMW lattice estimate, although at the cost of increasing the tension with the KLOE results.

\begin{figure}[t!]
    \centering
    \includegraphics[width=0.85\textwidth]{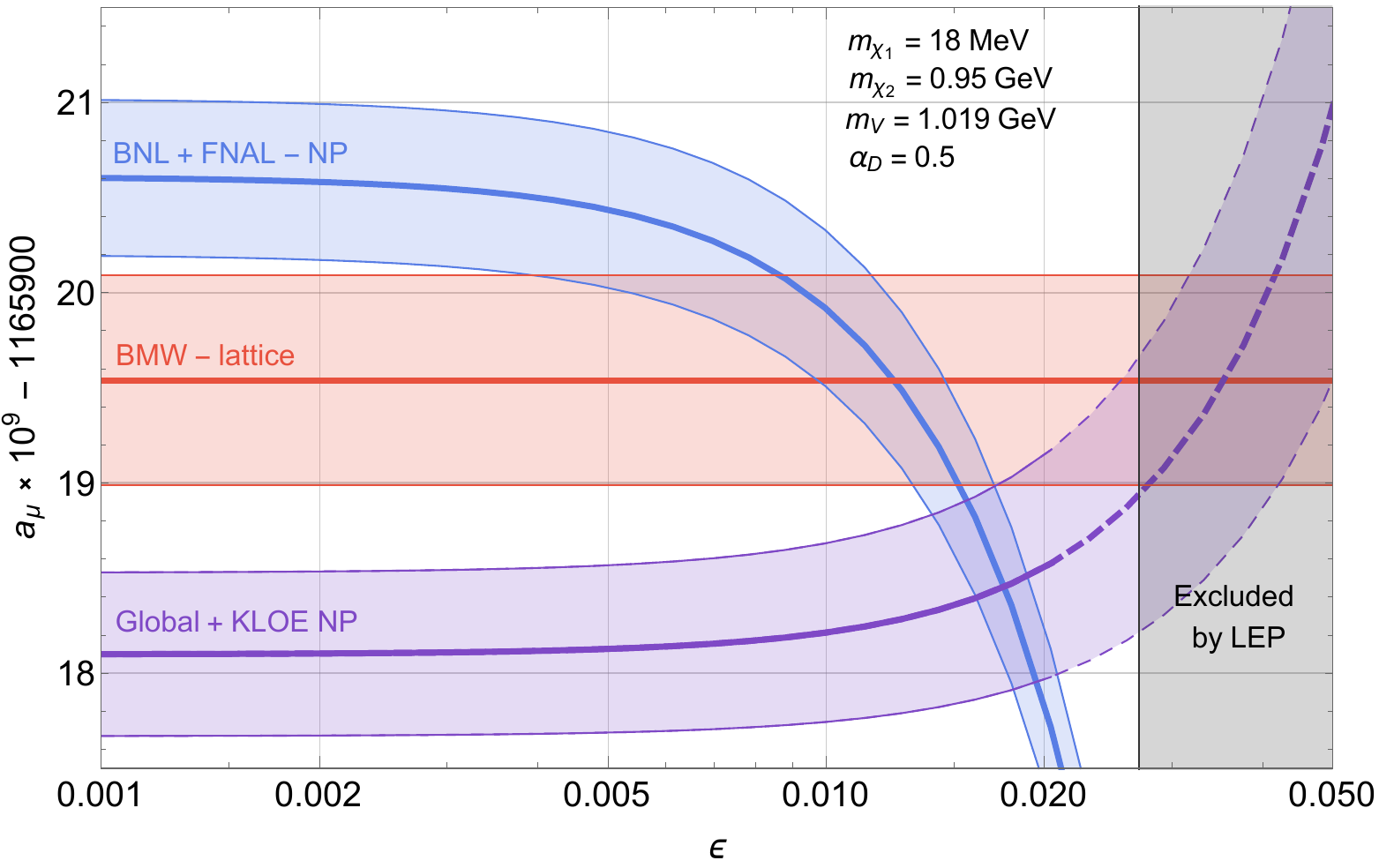}
    \caption{Theoretical prediction (purple) for $a_\mu$ as a function of $\varepsilon$ for a dark photon model with $m_{\chi_1} = 18$ MeV, $m_{\chi_2} = 0.9$ GeV, $m_V=1.019$ GeV and $\alpha_D=0.5$. The dashed purple curve denotes the region where the KLOE08 and KLOE10 results are more than $3\sigma$ away.
The blue band corresponds to the combined BNL and FNAL experimental results after subtracting the direct NP contribution from the dark photon. The red band shows the  prediction obtained with the BMW lattice estimate of $\amhvplo$. The  width of the bands represents $1\,\sigma$ uncertainties. The grey region is excluded by LEP. 
    \label{fig:final_plot}}
\end{figure}

In Fig. \ref{fig:final_plot} we show in purple the $\pm1\sigma$ band for the theoretical prediction of $a_\mu$ as a function of $\varepsilon$ for $m_{\chi_1}=18$ MeV, $m_{\chi_2} = 0.95$ GeV, $m_V = 1.019$ GeV and $\alpha_D = 0.5$. Since our analysis only includes the correction to $\amhvp$ in the $ \sqrt{s} \in [0.6, 0.9] \, \rm GeV$ range, we use a theoretical uncertainty on our result of $35\%$ to account for the missing contribution. The dashed curve denotes the values of $\varepsilon$ for which the KLOE08 result is more than $3\sigma$ away from  the KLOE10 measurement. 
The red region shows the $\pm1\sigma$ BMW-lattice computation and the blue region shows the $\pm1 \sigma$ band for the BNL and FNAL experimental results after subtracting the direct contribution to $a_\mu$. The grey area is excluded by a fit to the electroweak SM couplings at LEP, see eq. \eqref{eq:lep-ew-fit}. Therefore, this figure shows the interplay between the indirect effect of  correcting  the KLOE08, KLOE12 and BESIII results (purple curve) and the direct NP contribution to $a_\mu$ (blue curve), and identifies the regions where  the experimental result and the lattice and data-driven computation  become compatible at the $\sim2\sigma$ level. In particular the figure 
shows that with a specific  choice of the model parameters it is possible 
to bring the data-driven theoretical prediction in good agreement with the experimental result 
while reducing the  discrepancy with the prediction based on the lattice computation 
below the  $\sim 2\sigma$ level.

\section{Conclusions}
\label{sec:conclusions}
In this work, we have explored the intriguing possibility that a feebly interacting particle with mass around the $\phi$ resonance, that would be  
produced in  $e^+e^-$ collisions whenever $\sqrt{s}\geq m_\phi$, 
 could contribute  to the measured number of  $e^+e^-$   and $\mu^+ \mu^-$ events.
This could affect significantly   the 
estimated luminosity from Bhabha events, thus affecting the determination of 
$\shad$ from the number of hadronic final state. 
It would furthermore contribute to the ratio between ISR 
$\pi\pi\gamma$ and $\mu\mu\gamma$ events used in luminosity independent 
measurements of $\shad$, implying that some corrections must be applied before  
the experimental results could be related to the hadronic contribution to the 
photon vacuum polarization. 
Finally it would also require a reinterpretation of the   
results on measurements of the di-muon cross-section.
Such a cascade of rather subtle indirect effects can result in increasing the data-driven estimate of $\amhvp$ by a few percent, which  
suffices to solve the tension between the  KLOE and BaBar  determinations of $\shad$, 
and to reconcile the  estimate of the hadronic vacuum polarisation contribution $\amhvp$ from  
the data-driven dispersive method with the latest lattice calculation~\cite{Borsanyi:2020mff} 
(as well as with the estimate of $\amhvp$ from $\tau$ hadronic decays).

To illustrate the viability of this mechanism, 
we have constructed an explicit model based on a iDM paradigm which includes a  
vector particle mediator $V$ with  mass $m_V\sim 1\,$GeV. $V$  decays mainly proceed via a semi-visible leptonic channel, and can mimic effectively the Bhabha signature altering the 
estimated value of the luminosity at KLOE08, 
and produce the required excess of semi-visible muonic final states in the 
 KLOE12 and BESIII analyses. 
We have found that in our model the  indirect
effects  by themselves  are not sufficient to solve completely the discrepancy between the experimental determination  of $a_\mu$ and the prediction. However, the vector 
boson mediator will also contribute directly to 
$a_\mu$ via genuine NP loop contributions, and we 
find that when direct and indirect  effects are considered together, 
the $a_\mu$ discrepancy can indeed be solved, 
(with about $1/4$ of the discrepancy accounted by indirect effects, and  
$3/4$ by direct loop effects). 
The simple  model that we have put forth also  provides 
an adequate light dark matter candidate with a rich phenomenology which 
might be worthwhile exploring further.
On the long run,  we hope that our work can provide further motivations to the search for similar ``stealthy'' dark photons characterised by the required semi-visible 
decays.
Eventually, since the most important effect 
in our construction is that of shifting $\amhvp$ 
to larger values, a crucial test of the whole idea 
could come from the MuonE experiment \cite{Abbiendi:2016xup,Abbiendi:2020sxw}, as well as 
from new high precision determinations of $\amhvp$ on the lattice.

\bigskip
\paragraph{Note added} As this paper was being finalised, the work~\cite{DiLuzio:2021uty} appeared which 
studied the possibility of reconciling the data driven and the lattice determinations of $\amhvp$ by invoking a light NP mediator that directly modifies $\shad$. The authors  concluded that such a possibility is excluded by a number of experimental constraints. 
The NP effect explored in this work is of a completely different nature. As we have shown, it can not only reconcile 
the data driven and lattice determination of $\amhvp$, 
but it can also bring into agreement the KLOE and BaBar 
results for $\shad$.

\acknowledgments
We are indebted with G. Venanzoni for pointing out to us the KLOE12 luminosity 
independent determination of $\shad$~\cite{KLOE:2012anl} that triggered the revision 
of a first draft, and with S. M\"uller for valuable information on the KLOE12 analysis.
We thank M. Raggi and T. Spadaro for discussions,  and 
F. Piccinini and C.M. Carloni Calame for providing us with details 
on the BabaYaga event generator. LD thanks  Olcyr Sumensari and  Diego Guadagnoli for useful discussions. 
This work has received support by the INFN Iniziativa Specifica Theoretical
Astroparticle Physics (TAsP). G.G.d.C. is supported by the Frascati National 
Laboratories (LNF) through a Cabibbo Fellowship, call 2019.  This project has received funding from the European Union’s Horizon 2020 research and innovation programme under the Marie Sklodowska-Curie grant agreement No. 101028626.

\bibliographystyle{JHEP2}
\bibliography{bibliography}

\end{document}